\documentclass[fleqn,10pt]{wlscirep}
\usepackage[utf8]{inputenc}
\usepackage[T1]{fontenc}
\usepackage{moreverb, url}
\usepackage{soul}  
\usepackage{graphicx}
\usepackage{amsmath}
\usepackage{amssymb}
\usepackage{xcolor}
\usepackage{bbold}
\usepackage{url}
\usepackage{lineno}
\usepackage{rotating}

\title{Bayesian models for survival data of clinical trials: Comparison of implementations using R software}

\author[1,*]{Lucie Biard}
\author[1,2]{Anne Bergeron}
\author[1]{Sylvie Chevret}
\affil[1]{Université de Paris, INSERM UMR1153 - Team ECSTRRA,  AP-HP H\^opital Saint Louis, Paris, France}
\affil[2]{Service de Pneumologie, AP-HP H\^opital Saint Louis, Paris, France}

\affil[*]{lucie.biard@univ-paris-diderot.fr}

\date{18 August 2019}


\begin{abstract}
\textbf{Objective}: To provide guidance for the use of the main functions available in R for performing \emph{post hoc} Bayesian analysis of a randomized clinical trial with a survival endpoint using proportional hazard models. 

\textbf{Study Design and Setting}: Data derived from the ALLOZITHRO trial, conducted with 465 patients after allograft to prevent pulmonary complications and allocated between azithromycin and placebo; airflow decline--free survival at 2 years after randomization was the main endpoint.  

\textbf{Results}: Despite heterogeneity in modeling assumptions, in particular for the baseline hazard (parametric or nonparametric), and in estimation methods, Bayesian posterior mean hazard ratio (HR) estimates of azithromycin effect were close to those obtained by the maximum likelihood approach.

\textbf{Conclusion}: Bayesian models can be implemented using various R packages, providing results in close agreement with the maximum likelihood estimates. These models provide probabilistic statements that could not be obtained otherwise. 

\emph{Date: 18 August 2019}
\end{abstract}

\begin{document}

\flushbottom
\maketitle

\thispagestyle{empty}

\section{Introduction}
Bayesian approaches for clinical trials have received considerable interest in recent decades as a complementary tool to enhance the interpretation of clinical trials \cite{wijeysundera2009}. It has recently been highlighted \cite{lewis2018, ruberg_2019} that in the clinical trial setting, Bayesian analyses appear as a complementary tool for clinicians, patients, and policymakers.
However, Bayesian analysis of clinical trials is mostly restricted to the (often uncontrolled) phase II setting \cite{biswas2009} or phase III randomized clinical trials (RCTs) with a binary outcome measure \cite{goligher2018}. In the case of right-censored endpoints, a number of methods have been proposed for the Bayesian analysis of trial data by considering failure at a specific point in time and treating the data as binary. Indeed, compared to the straightforward beta-binomial framework for a binary variable, the use of Bayesian inference appears difficult to apply. This is notably true when the primary outcome is a survival outcome, and  although several papers and a book \cite{ibrahim2001} have been dedicated to Bayesian survival analyses, such methods are still underutilized in the clinical literature of RCTs \cite{brard2016}.

In such a comparative evaluation of survival, interest generally lies in the hazard ratios, which are classically estimated using the popular Cox regression analyses provided within standard statistical packages.
We thus focused our interest on the Cox proportional hazards (PH) model, and considered how, and with what functions from standard software, to implement a Bayesian estimation of the hazard ratio (HR) as the measure of the treatment benefit in an RCT.

Although many R packages are available for implementing survival models to handle right-censored data, only a few packages provide Bayesian estimation of those approaches. 
Moreover, all those packages differ in the modeling choice of the baseline survival, which complicates their comparison in practice. First, two main choices of modeling are considered, either fully parametric -- with survival times drawn from any parametric (exponential, piecewise constant hazard, etc.)  family -- or semi-parametric, with various nonparametric estimators of the survival functions. Second, these packages consider various modeling options for the prior density of the model parameters, including the regression coefficient measuring the randomization effect, and the additional baseline survival distribution that may include hyperparameters. Third, although model parameter estimation and statistical inference are mostly (though not consistently) conducted via self-tuning adaptive Markov chain Monte Carlo (MCMC) methods, the techniques may differ with respect to computational samplers, including starting values, algorithms and convergence diagnostics. 
Fourth, those packages are not easy to use in terms of defining all the parameters and priors. 
To improve the use and reporting of Bayesian analysis in survival trials as recommended \cite{rietbergen2017}, additional effort should be made to allow the appropriation of such methods by nonspecialized teams.

Therefore, the aim of this paper is to present and apply these different Bayesian parameterizations for PH survival models when analyzing data derived from an RCT with a survival outcome as the endpoint of interest. We also wish to assess whether those various estimates are relatively robust in terms of model conclusions for clinicians, patients, and policymakers. As an illustrative example, we use data from the ALLOZITHRO clinical trial (NCT01959100), that randomized 465 patients between azithromycin (234 patients) and placebo (231 patients) from February 2014 to August 2015, to prevent airflow decline after allogeneic stem cell transplantation for malignancy \cite{bergeron_2017}. This is a \emph{post hoc} Bayesian analysis of the main endpoint of the trial, airflow decline--free survival. For confidentiality reasons, we could not provide the original trial dataset; however, we used a resampled mock dataset that is similar to the data presented in the manuscript. 

For each of the reviewed PH models, we provide the results of their implementation using functions directly from R packages and data from the ALLOZITHRO example. 


\section{Bayesian models for survival data}

Let $N$ be the number of patients in the study, let $X_1,\ldots, X_N$, denote the right-censored time, and  let $Z_1,\ldots,Z_N$ denote a randomized group of individuals. In this paper, we consider the Cox proportional hazards model. 
Data are given by $(X_1, Z_1),\ldots, (X_N, Z_N)$, where :

\begin{equation}
\label{eq:ph}
X_i |Z_i \sim F(t|Z_i )=1-(1-F_0 (t))^{exp(Z_i' \beta)}, \ \ i=1, \dots, N
\end{equation}
Or, equivalently, 
\begin{equation}
H(t| Z_i )={exp(Z_i' \beta)} {H_0 (t)}
\end{equation}

\noindent where $F_0 (t)=1-S_0 (t)$ is the baseline distribution function of the right-censored time and $H(t| Z)= \int h(u|Z)du$  is the cumulative hazard function of the patients whose covariate is $Z$ and instantaneous hazard is $h_0(t)$.

\subsection{Partial Bayesian approach}\label{partialBayes}
In a RCT, given the main interest lies in the treatment effect, Bayesian inference frequently ignores the baseline hazard \cite{spiegelhalter2004}. 
This process results in modeling only the treatment effect size, which is measured by the log hazard ratio $\beta=\log{(HR)}$, considered to be normally distributed \cite{spiegelhalter2004}.

The normal prior for $\beta$ is expressed as $\mathcal N(\mu,\sigma_0^2)$, with $\sigma_0^2=4/n_0$ asymptotically, where $n_0$ is the "implicit" sample size of the prior information given by the effective number of events \cite{tsiatis1981, spiegelhalter1994, spiegelhalter2004}. This value is usually computed using the Schoenfeld formula as $n_0=\frac{(z_{1-\beta}+z_{1-\alpha/2})^2}{p^2\mu^2}$, where $\alpha$ is the type I error rate, $\beta$ is the type II error rate, $z_{1-x}$ is the $(1-x)$ quantile of the standard normal distribution, and $p=1/2$ is the proportion of patients allocated to each arm (balanced RCT). 

Given the normality of the likelihood estimate $y$ of the $\log{(HR)}$ on the data, with a sampling distribution $y\sim \mathcal N(\beta,\sigma^2)$, where $\sigma^2$ is the variance of the estimate, the computation of the posterior distribution is direct. Indeed, the application of Bayes' theorem gives the following posterior distribution:

$$\beta| y \sim \mathcal N \left[\frac{\mu\sigma^2+y\sigma_0^2}{\sigma^2+\sigma_0^2},\frac{\sigma^2\sigma_0^2}{\sigma^2+\sigma_0^2}\right].$$

The method first requires the specification of a mean prior for $\beta$, the $\log{(HR)}$; according to Spiegelhalter \emph{e al.}\cite{spiegelhalter1994}, two priors, a skeptical prior centered on the null hypothesis (that is, $\mu=0$) and an enthusiastic prior centered on the alternative hypothesis, are recommended.
Then, the posterior mean of $\beta| y$ is directly computed without the need for any specific function from the R package. 
The \emph{pnorm()} and \emph{pqnorm()} functions in R enable the posterior probabilities of specific intervals of $\log{(HR)}$ values and credibility intervals of the estimates, respectively, to be computed.

\subsection{Fully Bayesian approaches}

By contrast, in a fully Bayesian approach, the model in equation (\ref{eq:ph}) has two parameters, namely, $\beta$ and $F_0$, that require specification of priors. Although the specification of baseline survival can equivalently use the survival function ($S_0$) or hazard functions ($H_0$ or $h_0$), Bayesian inference methods differ in this component of the model, which is considered as either drawn from a parametric distribution or not. We will see that defining priors on such different quantities may yield differences in estimates.

Unfortunately, such models are not analytically tractable, and inference about parameters or quantities of interest requires approximate integration, such as Laplace's method, or an approximate technique, such as MCMC. In MCMC, samples can be generated from a posterior density and used to approximate expectations of the quantities of interest. These methods were first implemented in specific statistical software. Following WinBUGS in 1997 \cite{winbugs}, JAGS \cite{jags} and Stan \cite{stan} were developed to achieve this aim and can be freely downloaded. Recently developed \texttt{r2jags} and \texttt{rstan} R packages provide an interface for their use with R. Additionally, the \texttt{INLA} package implements integrated nested Laplace approximations for Bayesian inference as an alternative approach to MCMC \cite{inla}. Moreover, in the setting of Bayesian survival, several R libraries have been developed to address specific statistical aspects: for instance, \texttt{dynsurv} focuses on Cox models with time-varying or dynamic covariate coefficients for interval-censored and right-censored survival data \cite{wang2013}, \texttt{spBayesSurv} addresses spatial survival data \cite{zhou2018}, \texttt{survHE} \cite{survHE} proposes survival models for health economics studies, \texttt{rstanarm} constructs regression models using Stan including a joint survival model, and \texttt{SemiCompRisks} estimates hierarchical multistate models for the analysis of independent or clustered semicompeting risks data \cite{semicomprisks}.

Most of these models consider a normal prior for $\log{(HR)}$:

$$\beta \sim \mathcal N(\mu_0, \sigma_0^2).$$

Flat priors are commonly set as default in these packages, with mean $\mu_0=0$ and varying standard deviation $\sigma_0$, for instance, from 1 in the \texttt{dynsurv} package up to $10^{10}$ in the \texttt{spBayesSurv} package \cite{zhou2018}. 

The packages differ mostly in modeling the baseline survival function. Their main differences are detailed below and summarized in Table \ref{tab:diffPackages_2}.

\begin{sidewaystable}[htp!]
\begin{footnotesize}
    \centering
    \begin{tabular}{llc}
    \hline
      Package/Method  & Prior baseline survival or risk & Prior $\log{(HR)}$ \\
        \hline
        \multicolumn{3}{l}{\underline{\texttt{SemiCompRisks} package}}\\
        \emph{BayesSurv$\_$HReg(..., id=NULL, model="Weibull")}& $h_0(t)=\alpha \kappa t^{\alpha-1}$, $\alpha \sim \Gamma(a,b)$, $\kappa \sim \Gamma (c,d)$  &$\beta \propto 1$ \\
        \emph{BayesSurv$\_$HReg(..., id=NULL, model="PEM")} & $\log{(h_0(t))} = \sum_{k=1}^{K+1}{\log{(h_k)} I\{t \in \left[s_{k-1},s_{k}\right]\}}, K \sim Poisson(\alpha),$&$\beta \propto 1$\\
        &$\lambda|K, \mu_{\lambda}, \sigma_\lambda^2 \sim MVN_{K+1}(\mu_\lambda \mathbb{1}, \sigma_\lambda^2\Sigma_\lambda), s|K \propto \frac{(2K+1)!\prod_{k=1}^{K+1}{(s_k - s_{k-1})}}{(s_{K+1}) ^{(2K+1)}}$&\\
        &$\mu_\lambda \propto 1, \sigma_\lambda^{-2}\sim \Gamma(a,b)$&\\

\multicolumn{3}{l}{\underline{\texttt{INLA} package}}\\
 \emph{inla(inla.surv()}$\sim$\emph{.,..., family="exponential.surv")}&$h(t)=\lambda, \lambda=\exp{(1+\beta Z)}$ &$\mathcal{N}(0, 10^3)$ \\
  \emph{inla(inla.surv()~.,..., family="weibullsurv")}& $f(t)=\alpha t^{\alpha-1}\lambda \exp{(-\lambda t^\alpha)}$, where $\alpha >0, \lambda>0$, & $\mathcal{N}(0, 10^3)$ \\
  &shape $\alpha\sim \Gamma(25,25)$ and $\lambda=\exp{(1+\beta Z)}$&\\
\emph{inla(inla.surv()~.,..., family="coxph", control.hazard=list(model="rw1"))} &$\log{(h_0(t))}\sim$ Random Walk of order 1 by $M$ intervals ($M=15$) & $\mathcal{N}(0, 10^3)$\\
  & $\log{(h_{0,k})} - \log{(h_{0,k-1})}\sim \mathcal N(0, \sigma^2=\tau^{-1})$, with $\tau \sim \Gamma(1, 5.10^{-5})$  & \\
\emph{inla(inla.surv()~.,..., family="coxph", control.hazard=list(model="rw2"))} &$\log{(h_0(t))}\sim$ Random Walk of order 2 by $M$ intervals ($M=15$) & $\mathcal{N}(0, 10^3)$\\
  & $\left(\log{(h_{0,k})} - \log{(h_{0,k-1})}\right)^2 = \log{(h_{0,k})} - 2 \log{(h_{0,k+1})} + \log{(h_{0,k+2})}$ & \\
	& \quad \quad \quad $\sim \mathcal N(0, \sigma^2=\tau^{-1})$  with $\tau \sim \Gamma(1, 5.10^{-5})$&\\
  \multicolumn{3}{l}{\underline{\texttt{survHE} package}}\\      
\emph{fit.models(...,distr="exponential", method="hmc")}& $h(t)=\exp{(\beta_0 + \beta Z )}$ & $\mathcal{N}(0, 25)$ \\
\emph{fit.models(...,distr="weibull", method="hmc")}& $h(t)=\frac{\alpha}{\mu}[\frac{t}{\mu}]^{\alpha-1}$, with shape $\alpha \sim \Gamma(a=0.1,b=0.1)$,  &  $\mathcal{N}(0, 25)$\\
 &scale $\mu = \exp{(\beta_0+\beta Z)}$ & \\
  \multicolumn{3}{l}{\underline{\texttt{rstanarm} package}$^\star$}\\      
\emph{stan$\_$surv(...,basehaz="exp")}& $h(t)=\exp{(\alpha + \beta Z)}$, $\alpha \sim \mathcal N(0, 20^2)$  & $\mathcal{N}(0, 2.5^2)$\\
\emph{stan$\_$surv(...,basehaz="weibull")}& $h(t)=\exp{(\alpha + \beta Z)}$, $\alpha \sim \mathcal N(0, 20^2)$, shape$\sim \rm{Half~Normal}(0, 4)$ & $\mathcal{N}(0, 2.5^2)$\\
\emph{stan$\_$surv(...,basehaz="ms")}& spline coefficients $\sim \mathcal N(0, 20^2)$ & $\mathcal{N}(0, 2.5^2)$\\
\emph{stan$\_$surv(...,basehaz="bs")}& spline coefficients $\sim \mathcal N(0, 20^2)$  & $\mathcal{N}(0, 2.5^2)$\\
         
          \multicolumn{3}{l}{\underline{\texttt{dynsurv} package}}\\
          \emph{bayesCox(..., model="TimeIndep") } & $h_0(t) = \sum_{k=1}^M{h_k I\{t \in I_k\}}, \ h_k \sim \Gamma({\rm shape} = 0.1, {\rm rate} = 0.1)$&$\mathcal N(0, 1)$ \\
           \quad Piecewise constant &Intervals $I_k$ defined after the observed finite censoring endpoints& \\
%
          \multicolumn{3}{l}{\underline{\texttt{spBayesSurv} package}}\\
        \emph{indeptCoxph()}& $h_{0k}=\Gamma(r_0h, r_0)$, with $r_0=1, h = \hat{h}, k=1, \dots, M$   &$\mathcal N(0, 10^5)$\\
         \quad Piecewise constant& $M=10$ ($d_k$=deciles) &\\

         \emph{survregbayes()}& $S_0(.)|\alpha, \theta \sim TBP_L(\alpha, S_\theta(.)), \ \ L=15$&$\mathcal N(0, 10^{10})$\\
          \quad Transformed Bernstein Polynomial   & $\alpha \sim \Gamma (a_0, b_0)$, with $a_0=b_0=1$ &\\
 \quad (TBP) & $S_\theta(t|\theta_1, \theta_2) = 1 - \exp{\{ -(e^{\theta_1}t)^{\exp{(\theta_2)}} \}} $ (Weibull) & \\
& or $S_\theta(t)=\{1+(e^{\theta_1}t)^{\exp{(\theta_2)}}\}^{-1}$(log-logistic)&\\       
& or $S_\theta(t)=1 - \Phi\{(\log{t+\theta_1})\exp{(\theta_2)}\}$ (log-normal)&\\
& $(\theta_1, \theta_2)=\mathcal N (\boldsymbol{\theta_0}, \boldsymbol{V_0})$&\\
& $\boldsymbol{\theta_0}= \boldsymbol{\hat{\theta}}, \boldsymbol{V_0}=\boldsymbol{\hat{V}}$, the ML estimates with the parametric model& \\

         \emph{survregbayes2()}& Centering distribution: Weibull, log-logistic, log-normal  & $\mathcal N(0, 10^{10})$\\
               & \quad (see \emph{survregbayes()} above)&\\
								\quad Mixture of Polya tree (MPT) & Default Polya tree level = 4&\\
\multicolumn{3}{l}{\underline{Murray \emph{et al.} (2016) function}}\\
        PH model & $\log{(h(t; \alpha^\star))}=\alpha_0^\star + \alpha_1^\star + \sum_{k=2}^{K}{\alpha_k^\star(|t-\Tilde{t}_{k-1}| - |\Tilde{t}_{k-1}|)}$, & $ \mathcal N(0, 10^{4})$\\
        & $k=2, \dots, K, \ \ K=20$ & \\
				& $\alpha_0 \sim \mathcal N(0,10^4), \alpha_1 \sim  \mathcal N (0,10^4), \alpha_k|\sigma_{\alpha} \sim \mathcal N(0, \sigma_\alpha^2)), \ \ \sigma_\alpha \sim \mathcal U(0.01, 100)$ &\\  
				& See transformation from $\alpha^\star$ to $\alpha$ in Murray et al (2016) & \\
         
         \hline
    \end{tabular}
    \caption{Model default parameterization in R packages for Bayesian approaches}
    \label{tab:diffPackages_2}
\end{footnotesize}
{\footnotesize $^\star$ As of 22 July 2019, \texttt{rstanarm} development version, including the \emph{stan$\_$surv()} function, downloaded from the survival branch of the package available on github at https://github.com/stan-dev/rstanarm/tree/feature/survival}
\end{sidewaystable}

\subsubsection{Parametric PH models} \label{sec:parametric}
\underline{Exponential and Weibull models} -
The simplest way to provide Bayesian inference in a survival model is to assume the time to event is drawn from a parametric distribution. The models have the same multiplicative structure as the Cox PH model, but rather than leaving the baseline hazard unspecified, it is assumed to be a power transform of the time scale. This model formulation allows for the fact that the baseline hazard over time may be constant (exponential distribution), increasing or decreasing (Weibull distribution).
The use of a parametric baseline survival results in a fully parametric PH model. 
This method was used for empirical Bayesian analysis by Kalbfleish \cite{kalbfleish1978}, with the conclusion of avoiding the assessment of data by using only one parametric survival model \cite{omurlu2015}. 

The likelihood for such models can be shown to have the following form
$$L(\theta,\beta)=\prod_{i=1}^{n} \left(h_0(t|\theta)e^{\beta^TZ_i}\right)^{\delta_i} exp\left[-\int_0^{t_i} h_0(u|\theta)e^{\beta^TZ_i} du\right]$$
where $h_0(t|\theta)$ is the baseline hazard function, dependent on a vector of unknown parameters $\theta$, $\delta_i$ is an indicator of failure for the $ith$ patient, taking a value of one if the $ith$ patient experienced failure and zero otherwise, and $t_i$ is the time of failure or censoring time for the $ith$ patient. 

Bayesian inference incorporates prior information expressed as a probability density function on so-called hyperparameter(s)  $\theta$ of the survival distribution, that requires estimation
\cite{abrams1996}. The priors for the hyperparameters are commonly based on log-normal or gamma distributions.
The prior information on $\theta$ and $\beta$ can then be combined with the observed data likelihood described above, using Bayes' theorem, to obtain a posterior density function.

Such models are not analytically tractable, and inference about the parameters of interest requires approximation methods or MCMC, implemented using BUGS, JAGS or Stan, for instance. Recently, R packages specifically dedicated to survival data have been developed \cite{brard2016, omurlu2009}. Thus, models are currently readily available in the \texttt{SemiCompRisks} package (function \emph{BayesSurv$\_$HReg(, model="Weibull")}) \cite{semicomprisks}, in the \texttt{INLA} package \cite{inla}, in the \texttt{rstanarm} package (function \emph{stan$\_$surv()} with argument \emph{basehaz="exp"} or \emph{basehaz="weibull"}), and in the \texttt{survHE} package (which allows inference through either INLA via a call to the \texttt{INLA} package or Hamiltonian Monte Carlo simulation (HMC) via a call to \texttt{rstan}) \cite{survHE}. 

\underline{Piecewise exponential models} - To allow the baseline hazard to be more flexible, a piecewise exponential model, sometimes referred to as piecewise constant model, has been proposed, where the hazard changes from time to time during the observation period but remains constant within each time interval. 
This model thus requires the definition of a partition of $M$ intervals $I_k= (d_{k-1}, d_k], k=1, \dots, M, d_0=0$ and $d_M=\infty$, where the baseline hazard is constant and equal to an unknown value, $h_k>0$: 
$h_0(t)=\sum_{k=1}^M{h_k I\{t \in I_k\}}$. The parameters of interest are then $h_k$, which are associated with priors, usually from gamma distributions. 

In the most basic form, a piecewise constant hazard model can be obtained via the so-called "Poisson trick". Provided a reformatting of the time-to-event data aggregated by time intervals (number of events and total patient-time available for each time interval), a Poisson regression model is used to estimate the treatment effect and $K$ independent piecewise constant hazards $h_k$. A dependence among piecewise hazards may also be defined in dedicated models. 

PEM models are for instance implemented in the \emph{indeptCoxph()} function of the R \texttt{spBayesSurv} package \cite{zhou2018}, 
where $d_k$ is set as the $\frac{k}{m}$th quantile of the empirical distribution of the observed survival times for $k=1, \dots, M-1$:
$$h_k \sim \Gamma(r_0h, r_0)$$ 
with $h_k$ assumed to be independent, and the default hyperparameter values are set to $M=10, r_0~=~1, h~=~\hat{h}$, the maximum likelihood estimate of the rate in an exponential PH model. 
Another implementation is available in the \texttt{SemiCompRisks} package, with the function \emph{BayesSurv$\_$HReg()}, 
with the interval partition sampled from a time-homogeneous Poisson process prior (with multinormal prior on the piecewise log-hazards, centered on zero and Gamma hyperprior, $\Gamma(a,b)$, on the variance terms, a Poisson prior on the number of intervals $M$, with hyperparameter $\alpha$ to be specified by the user).
Similarly, in the \texttt{INLA} package, PEM can be estimated with $\log{(h_0(t))}$ modeled over $M$ intervals with a random walk process of order 1 or 2 on the difference $\log{(h_k)} - \log{(h_{k-1})}$.

\subsubsection{Nonparametric priors}
A variety of Bayesian nonparametric priors have been applied to baseline survival distributions $S_0(t)$, which is straightforward since each PH model has a simple formula that relates survival to covariate information through this baseline and a linear function of the covariates. Those priors proposed in R packages are briefly described below.

\subsubsection*{Polya tree priors}
Fully Bayesian semiparametric survival models based on MCMC methods were developed by Walker \emph{et al. }\cite{walker1999} using Polya tree priors. More recently, mixtures of Polya tree (MPT) priors were also developed, in the setting of spatial frailty modelling \cite{hanson2006,zhao_2009, zhao_2011}. 
These methods are implemented in the function \emph{survregbayes2()} of the R \texttt{spBayesSurv} package \cite{zhou2018}, which notably allows for PH models without frailty (in case of non spatial data). Options for the centering distribution for the MPT prior include Weibull distribution, log-logistic and log-normal distributions (see table \ref{tab:diffPackages_2}). 

\subsubsection*{Transformed Bernstein polynomial}
The transformed Bernstein polynomial (TBP) prior has also been proposed for fitting the baseline survival function.
Unlike the mixture of Polya trees, the TBP prior selects smooth densities, leading to efficient posterior sampling. Given a fixed positive integer $L$, a TBP prior follows:

$$S_{0}(t)=\sum_{j=1}^{L}w_j I(S_{\theta}(t)|j,L-j+1),w_L\sim \rm{Dirichlet}(\alpha,\ldots,\alpha)$$

\noindent where $w_L=(w_1,\ldots,w_L)^T$ are positive weights, $I(.|a,b)$ is the beta cumulative function with parameters $(a,b)$, and $S_{\theta}$ is a parametric survival function with parameter(s) $\theta$ (\textit{e.g.}, log-logistic, Weibull, log-normal). 
The parameter $\alpha$ controls how close the shape of $S_0(t)$ is relative to the prior guess $S_{\theta}$: small values of $\alpha$ allow more pronounced deviations of $S_0(t)$ from $S_{\theta}$.
Thus, such Bernstein polynomials enable density estimation on bounded domains, typically by transforming data to lie in the interval $[0, 1]$. This process provides a convenient way to create a Bayesian nonparametric prior for smooth densities, leading to efficient posterior sampling.
This technique is implemented in the function \emph{survregbayes()} of the R \texttt{spBayesSurv} package \cite{zhou2018}.

\subsubsection*{Splines}
\underline{Piecewise log-hazard with low-rank thin plate (LRTP) splines}. Murray \emph{et al.} proposed a flexible framework using LRTP linear splines for a continuous log-hazard function, as opposed to the discontinuous hazard in piecewise exponential models described above \cite{murray_2016}. Of note, this method was developed specifically to handle time-dependent effects of covariates, but the model can be simplified for the assumption of proportional hazards across treatment groups. R functions are available online, calling JAGS for MCMC estimation (https://www.counterpointstat.com/software-and-peer-reviewed-literature.html). 

\noindent \underline{Cubic M- and B-splines} can be used to model the baseline hazard \cite{royston_2002}. Such models are to be available in a new survival branch of the \texttt{rstanarm} package, currently under development; for the present work, we used its development version, as available on Github. In absence of time-dependent effects, the estimation process is faster with M-splines than B-splines given a closed form for cumulative hazard is available only in this case.

Of note, piecewise spline modeling of the log-hazard is sometimes categorized along with PEMs; there are similarities in the modelling approach with partitioning of the time axis,  although, in the case of splines, the log-hazard is not piecewise constant contrary to the models described in section \ref{sec:parametric}. \\

For the methods using MCMC, the functions and packages also differ in terms of the default simulation parameters (number of iterations, chains, etc.). A description of the default parameterizations is available in Table S1 in the Supplementary material.

\section{Illustrative example derived from the ALLOZITHRO trial}
An example dataset was generated from the original ALLOZITHRO trial data: we randomly drew a sample with the same number of observations, using bootstrap (with replacement), stratified on the randomization arm.
At the time of analysis, 231 patients (50\%) had experienced a primary event: 129 in the azithromycin group and 102 in the placebo group. There was no evidence of any significant violation of the proportional hazards assumption, as assessed by the Grambsch and Therneau test ($P=0.60$) \cite{grambsch_1994}.

\begin{figure}[htp!]
   \centering
   \includegraphics[width=3.15in]{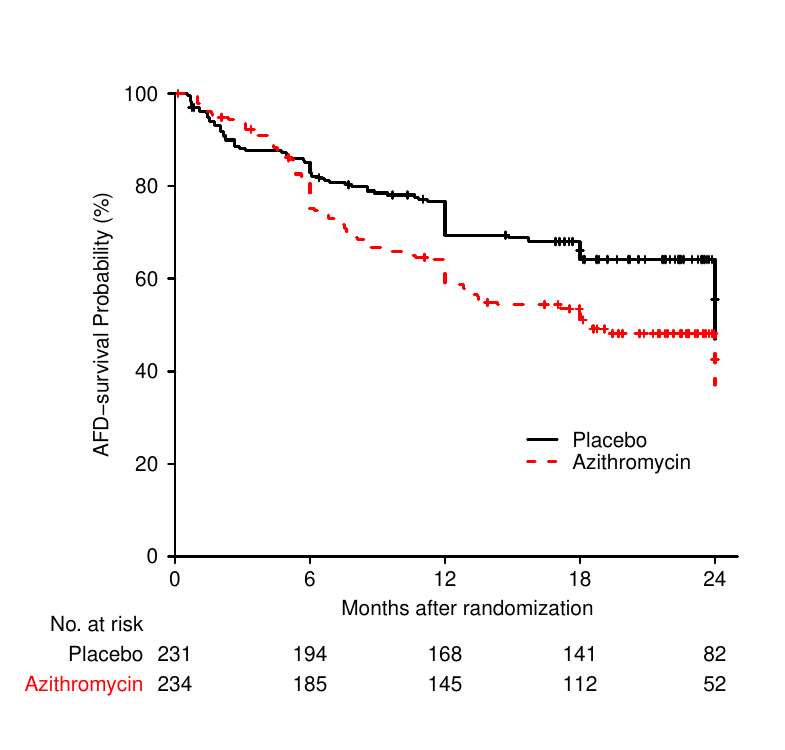}
   \caption{Estimated Kaplan-Meier AFD-survival according 
 to randomization}
   \label{fig:km}
\end{figure}

Using a standard data analysis based on maximum likelihood estimation of a semiparametric Cox PH model, the estimated hazard ratio of airflow decline (AFD)-free survival was $HR=1.44$, 95\%CI: 1.11 to 1.87 (p-value=0.006) 
(figure \ref{fig:km}).

We then applied the Bayesian approaches described above. 
Estimates of the posterior $\log{(HR)}$ throughout the different approaches, with their default parameterization (see details in Supplementary material, table S1), are summarized in Figure \ref{fig:forestplot} and Table S2 in Supplementary material. 

\begin{figure}[htp!]
    \centering
    \includegraphics[width=7in]{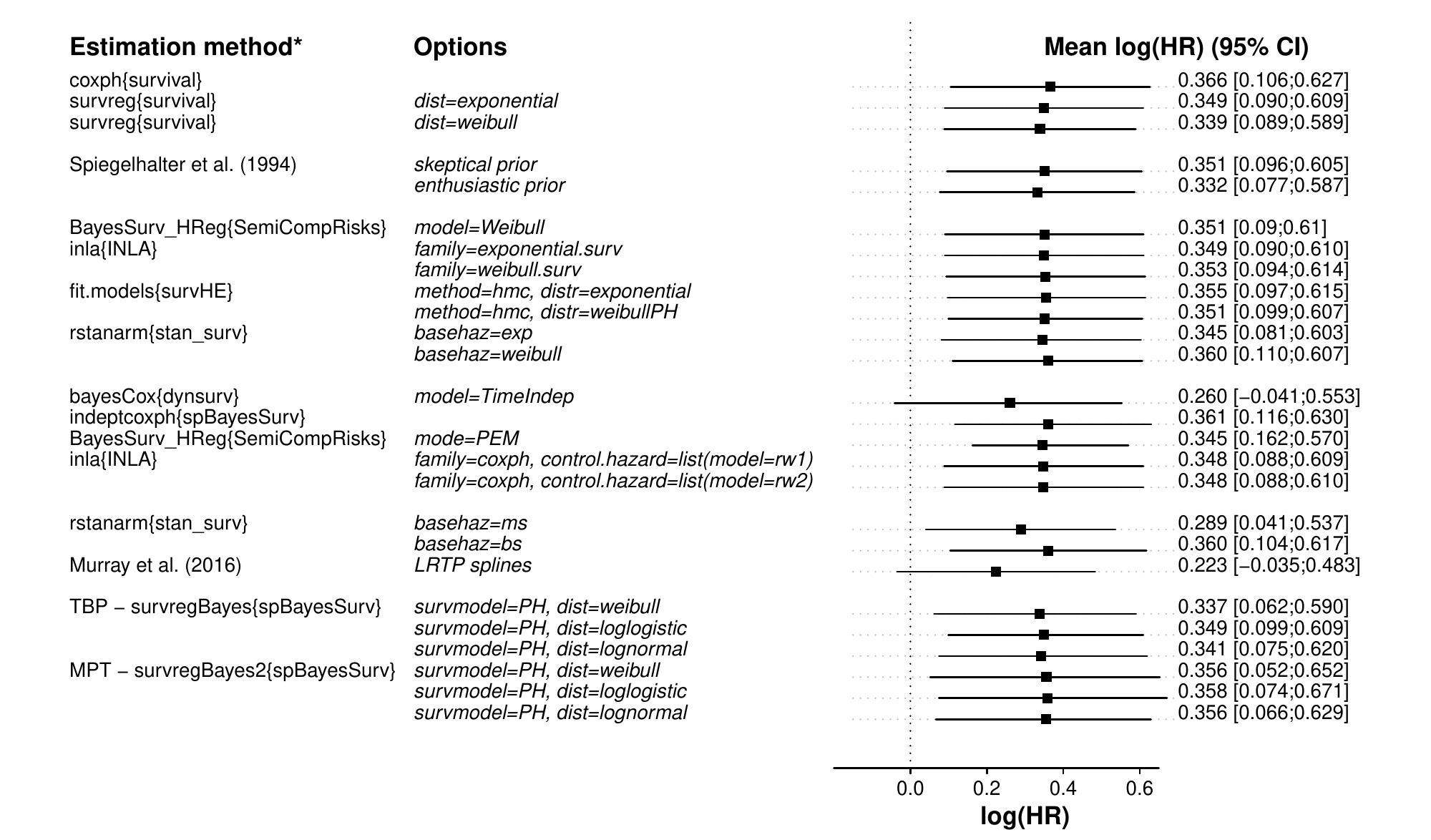}
    \caption{ALLOZITHRO Trial: Forest plot of $\log{(HR)}$ posterior estimates by type of model and R function, using default parameterization. Black squares represent the posterior mean of $\log{(HR)}$ and segments the 95\% credibility interval for Bayesian methods, the maximum likelihood point estimate and 95\% confidence interval for frequentist methods. $^\star$ function\{package\} or reference publication}
    \label{fig:forestplot}
\end{figure}

\begin{figure}[htp!]
    \centering
    \includegraphics[width =6.3in]{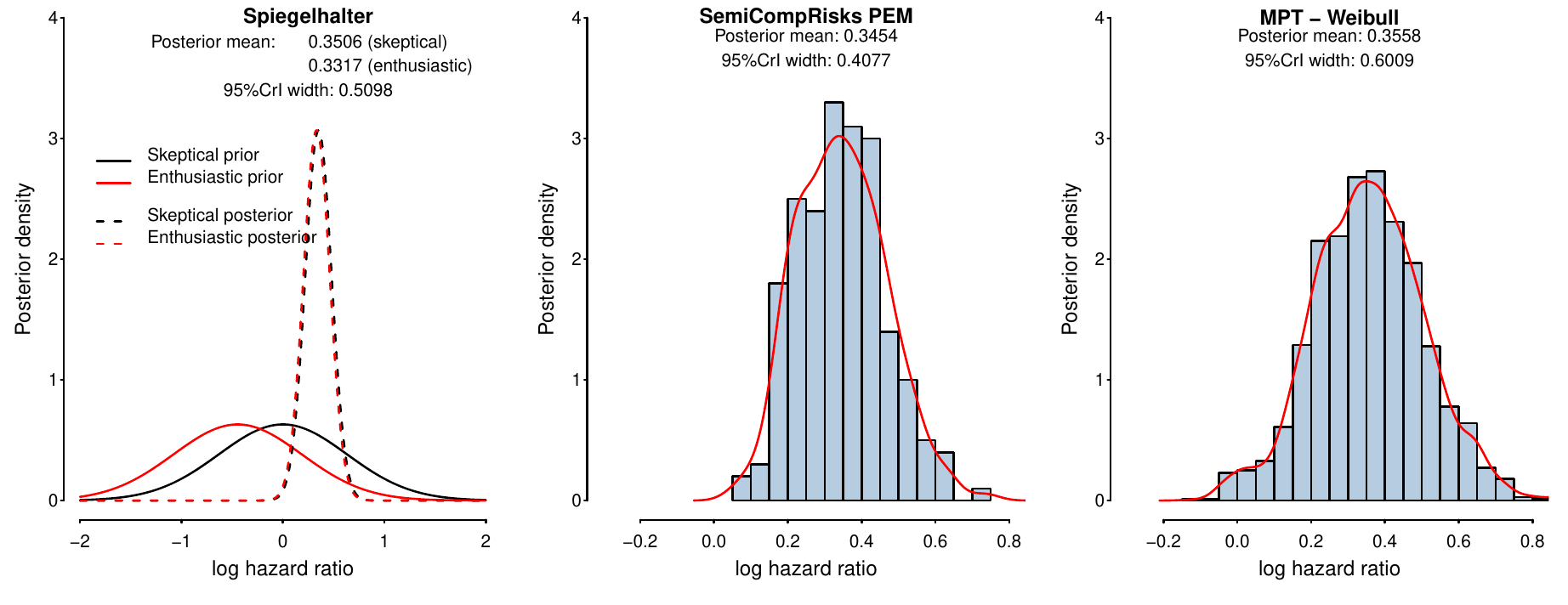}
    \caption{Bayesian posterior distribution of the log hazard ratio ($\log{(HR)}$) - Left panel refers to the non-fully Bayesian approach with skeptical or enthusiastic priors for log HR. The middle and right panels refer to fully Bayesian semiparametric PH models, with normal prior for $\log{(HR)}$ and either an exponential distribution, estimated with Stan-Hamiltonian Monte Carlo approach via the package \texttt{survHE} (middle plot) or a mixture of Polya trees (MPT) prior with centering Weibull distribution, estimated with the package \texttt{spBayesSurv} model (right plot) for the baseline survival function}
    \label{fig:spieg}
\end{figure}

\subsection{Non-fully Bayesian analyses}
We first applied the partial likelihood approach and model described in section \ref{partialBayes}. We considered normal priors for $\log{(HR)}$. As previously done \cite{moatti2013}, the skeptical prior corresponded to the belief that
there was no difference between groups, which was expressed by a prior mean for $\log{(HR)}$: $\mu = 0$, while the enthusiastic prior was set to $\mu=log(0.64)=-0.45$, corresponding to the postulated expected 15\% benefit over the 2 year estimated AFD-free survival of 45\% in the control arm, which was used in the protocol to compute sample size. Other prior parameters were set to $n_0=10$ and $\sigma_0=0.63$, using the formulas described above. 

On the basis of the Bayesian partial conjugated normal model, $\log{(HR)}$ was set to $y=0.366$ with a standard error of $\sigma=0.133$. This resulted in a posterior mean $\log{(HR)}$ of 0.351 or 0.332 for the skeptical or enthusiastic prior, respectively (figure \ref{fig:spieg}, left panel). Compared to the standard HR estimate of 1.44, these Bayesian analyses produced similar mean posterior skeptical and enthusiastic estimates of HR of 1.42 and 1.39, respectively. Bayesian inference also provided a posterior probability of $HR>1.5$, which was computed for both priors, skeptical and enthusiastic, at 0.34, and 0.29, respectively.

\subsection{Fully Bayesian analyses}

We then applied parametric and semi-parametric fully Bayesian survival models. MCMC convergence diagnostics are available in Supplementary material (Figures S1 and S2). 

\subsubsection*{Parametric baseline hazard}
We first considered exponential and Weibull models for AFD-free survival, using different packages for comparison purposes, with their default parametrization for MCMC: \texttt{SemiCompRisks}, \texttt{INLA},  \texttt{rstanarm}, \texttt{survHE}. The default priors on $\log{(HR)}$ were flat noninformative: uniform with mean zero for \texttt{SemiCompRisks} and $\mathcal N(0,10^3)$ for \texttt{INLA} and \texttt{survHE}, or weakly informative centered on zero for \texttt{rstanarm}, with $\mathcal N(0,2.5^2)$.
The mean posterior $\log{(HR)}$ ranged from 0.345 to 0.360, which was close to the maximum likelihood estimates for these two parametric models (figure \ref{fig:forestplot}).

We then applied piecewise exponential models for the baseline hazard by using the following functions: \emph{bayesCox()} of the \texttt{dynsurv} package, \emph{indepdtCoxph()} of the \texttt{spBayesSurv} package, \emph{BayesSurv$\_$HReg()} of the \texttt{SemiCompRisks} package and \emph{inla.coxph()} of the \texttt{INLA} package. Similarly to the previous models, noninformative flat priors for $\log{(HR)}$ centered on 0 with very large variances resulted in posterior means close to the maximum likelihood estimates in the Cox and parametric models, suggesting that the prior was swamped with the data. An exception was \emph{bayesCox()} in \texttt{dynsurv}, which yielded a $\log{(HR)}$ of 0.260, with 95\% credibility interval (-0.042; 0.553).

\subsubsection*{Nonparametric baseline hazard}
The results from models with nonparametric estimation of the baseline hazard were consistent whether the prior was defined by a transformed Bernstein polynomial (TBP) or a mixture of Polya trees (MPT) and regardless of the underlying centering density family, with the exception of the function provided by Murray \emph{et al.}, with LRTP splines for a piecewise linear log hazard (mean posterior $\log{(HR)}$= 0.223) \cite{murray_2016} and the model with M-splines for the log hazard in \texttt{rstanarm}, with $\log{(HR)}=0.289$ (95\%CrI 0.041; 0.537) (Figure \ref{fig:forestplot}). 

Note that the posterior variance was somewhat lower with the partial Bayesian models and the exponential model implemented in the package \texttt{survHE} with Stan, as illustrated in figure \ref{fig:spieg}.

\subsubsection*{Sensitivity analyses}
Secondarily, we re-estimated the models requiring MCMC using homogenized MCMC parameters for the number of chains and optimized MCMC convergence (total number of iterations, thinning) (Table S3). Results are available in the Supplementary material; estimates of the $\log{(HR)}$ were consistent with those in the main analysis and are reported in Table S4, Figure S5.

\section{Discussion}
Bayesian inference has been proposed as an alternative way for clinicians and policymakers to analyze data from RCTs \cite{ruberg_2019}. Indeed, with the development of Markov chain Monte Carlo (MCMC) methods and ever-growing computational capacities, it has become possible to
fit increasingly large classes of models with the aim of exploring the real-world complexities of
data. This could apply to survival models for right-censored endpoints, which are widely used in cancer RCTs. However, only a few works have been published on Bayesian survival methods. Moreover, they all use various model formulations, various priors, and integration methods that make their understanding and implementation difficult. We thus aimed to provide a review and practical guidance for using the main approaches, in open source R software platform.

Except for the \emph{bayesCox()} function of the \texttt{dynsurv} package and the model proposed by Murray \emph{et al}\cite{murray_2016} with LRTP splines for the piecewise linear baseline hazard function, all posterior mean estimates of azithromycin effect were close on the $HR$ scale based on either parametric or nonparametric modeling of the baseline hazard or the non-fully Bayesian approach. These results were obtained using the default parameterizations of the packages, ranging from 1.39 (with the non-fully Bayesian method, with the enthusiastic prior) up to 1.43 (with the \emph{indeptCoxph()} function of the \texttt{spBayesSurv} package). Nonparametric estimates based on transformed Bernstein polynomials or mixture of Polya trees priors for the baseline survival function were between those estimates. Moreover, the upper Bayesian estimate of 1.43 was close to that obtained by the maximum likelihood (that is, $HR=1.44$). 

Estimates of $\log{(HR)}$ were not markedly affected by increasing the variance of the prior (from 1 up to $10^{10}$). This results suggests that some one- or two-parameter baseline survival models should be avoided in this setting. 

In all cases, the Bayesian approach further allowed the computation of posterior probability intervals, such as $Pr(HR>1.5)$, which cannot be obtained by the standard frequentist approach.

In the piecewise modeling of the time to event functions (either hazard or log-hazard), the choice of partition of the time domain appears to be crucial. The \emph{bayesCox()} function of the \texttt{dynsurv} package defines, by default, the time intervals by the observed failure times in the dataset, with the direct consequence, in our case, of a larger number of intervals and longer computation time, as advised by the package's authors for large datasets \cite{wang2013}. Moreover, the obtained mean $\log{(HR)}$ estimate was far (HR=1.11) from that of the other methods. A reasonable approach requires enough time intervals $M$ to capture the curvature of the hazard function \cite{wang2013, murray_2016}.  In the setting of their flexible LRTP spline model of the log-hazard, Murray et al. argues for equally spaced partitions with a fixed number of $M$ time intervals when conducting simulations and using grid search over $M$ when analyzing a particular dataset to identify the partition that maximizes the criterion $deviance  + M$ \cite{murray_2016, spiegelhalter_2002}. Quantile-based or equally spaced partitions are common in the packages we described. When applied to the \emph{bayesCox()} function, instead of the default failure times approach, this practice yielded an estimate closer to that of the other methods, and in shorter time (posterior mean $\log{(HR)}=$0.4058 [95\% CrI: 0.1223 to 0.6919], 0.4004 [95\%CrI: 0.1102 to 0.6987] with 5th-percentile partition or 20 equally spaced intervals on the observed failure times, respectively). Such an impact of the parameterization of the time partition, resulting in the $\log{(HR)}$ estimate increasing from 0.2604 to 0.4058 in this example, should be emphasized.

Another practical issue that was encountered was the need to rescale the observed failure times for some models to fit. This requirement is expected for the Weibull model using the \texttt{INLA} package to avoid numerical overflow in the likelihood routine due to large time values \cite{inla}. For instance, with our dataset, the event times ranged from 0.43 to 24 months; we therefore used the time expressed in years instead of months to run this model. The same effect was observed with the function of Murray \emph{et al}\cite{murray_2016}.

In addition to differences in modeling assumptions, the packages also differ in the MCMC default parameterization (number of iterations, burn-in, number of chains). In some functions, the default number of iterations is small (2000) to avoid lengthy computations by default when initially assessing functions and models. However, more iterations are likely needed to fit models and obtain MCMC convergence. Note that computation time was somewhat increased with the \texttt{dynsurv} R package, when used with the default time intervals parameterization, as mentioned above. Similarly, some functions run one MCMC chain only for the posterior distribution estimation. Most functions rely therefore on user expertise to tune the estimation parameters and conduct diagnostics on MCMC estimation in a given modeling situation. For instance, MCMC mixing for Polya tree models has been reported to be problematic when the true baseline survival function is very different from the parametric family that centers the Polya tree and to require particular caution. In our example, this did not appear to be an issue. As with any statistical inference method, Bayesian model estimation requires a minimum of critical approach, and hands-on or turn-key use of default functions should be considered with caution to avoid erroneous conclusions. Workflow guidance and tools have been proposed to that aim \cite{gabry_2019}: for instance (not exhaustive), R packages such as \texttt{boa}, \texttt{coda}, \texttt{bayesplot} and \texttt{shinystan} have been developed to explore MCMC outputs \cite{boa, coda, gabry_2019}. Overall, the \texttt{rstan} outputs (used in the present analyses via the \texttt{survHE} and \texttt{rstanarm} packages) were the most comprehensive and straightforward in providing convergence information and diagnostics.

More complex parametric survival models have been implemented in R.
For instance, in the context of so-called general interval-censored data, a mixture of left-, right-, and interval-censored data, Lin \emph{et al.} used monotone splines to model the baseline cumulative hazard function \cite{lin_2015}. This approach is available in the \emph{ICBayes()} function of the R \texttt{ICBayes} package \cite{icbayes}; however, this method is not directly applicable to the only right-censored data typically observed in clinical trials. 
Otherwise, as proposed by de Iorio \emph{et al}, the survival function can be modeled nonparametrically by a mixture of Dirichlet processes (DPs)\cite{deiorio2009}. 
DP mixtures of kernels have been implemented in the function \emph{anovaDDP()} of the \texttt{spBayesSurv} package, with the choice of log-normal, log-logistic or Weibull kernels \cite{zhou2018}. More recently, Xu \emph{et al} proposed a DDP model with a normal kernel, the DDP-Gaussian Process model (DDP-GP), to model the log-failure time\cite{xu2018}. This method is implemented in the \texttt{DDPGPSurv} package \cite{ddpgpsurv}. Morevover, the DDP-GP model allows for individual scaling of included covariates, resulting in greater flexibility and robustness \cite{xu2018}. Imprecise or near-ignorance DP are used to estimate the survival function from right-censored data in the \texttt{IDPSurvival} package, function \emph{isurvfit()} \cite{mangili_2015}. The package provides a test to compare survival functions between groups, similar to a rank sum test (function \emph{isurvdiff()}). 
Note that in these flexible nonparametric modeling approaches, there is no necessity for resulting survival curve estimates to satisfy the ubiquitous proportional hazards assumption. Nevertheless, although they estimate the survival distribution according to covariate-defined groups, these approaches do not enable direct estimation of the average difference between groups, such as the HR.

We considered the proportional hazards setting of the Cox model, as it is the most commonly used survival model for randomized trials. The PH assumption was not considered violated in our example according to the Grambsch and Therneau test. Nevertheless, as illustrated in figure \ref{fig:km}, some departure from the assumption could be suspected. Nonproportional hazards models have been proposed in the Bayesian framework and could be of interest in such situations, in particular, in the development of immunotherapies \cite{alexander_2018}. For instance, the package \texttt{dynsurv} has been specifically developed for estimating time-varying effects of covariates using piecewise models; similarly, the method proposed by Murray et al. includes a model with a piecewise linear LRTP spline function for a time-varying hazard ratio \cite{murray_2016}. In the setting of joint models for prognostic studies with biomarkers, the R package \texttt{rstanarm} includes a model with a Weibull distribution for the hazard function and B-splines for the log hazard (function \emph{stan$\_$jm()}) \cite{peltola_2014, brilleman_2018_talk, rstanarm}). Similar models for health economics data were implemented with the R package \texttt{survHE}, with an underlying parametric baseline hazard function.

\subsection*{Limitations}
This study has some limitations. 
First, we used data derived from the ALLOZITHRO trial as a case study, and results may be dependent on the dataset. Second, we reported only a few posterior estimates, namely, the posterior mean of the $\log{(HR)}$ with the 95\% credibility interval and the posterior probability of the $\log{(HR)}$ being greater than 1.5, while other probabilistic statements could have been used to further describe the posterior knowledge on the treatment effect. Indeed, we believe that providing multiple posterior estimates of the parameter of interest is of prime interest in Bayesian analyses.
Third, we focused on the R software platform to review available functions and libraries. For the \texttt{rstanarm} package, we used the development version currently available for installation from the survival branch of the package on Github platform, to access the \emph{stan$\_$surv()} function. It will be integrated in the next version of \texttt{rstanarm} to be available for standard download. Survival models are also implemented in other software, such as SAS \cite{sas}, Stata \cite{stata}, and Python with the \texttt{survivalstan} library. Nevertheless, R statistical software is an open-source platform that allows centralizing various models and estimation algorithms (JAGS, Stan, custom functions), conveniently enabling the comparison of methods from the same console.

\subsection*{Conclusion}
In summary, we follow the statement that 'all models are wrong, but some are useful', the aphorism generally attributed to the statistician George Box. When applied to survival analysis, we showed that Bayesian models could be implemented using R packages, providing results generally in agreement with the likelihood estimates. The non-fully Bayesian approach with a conjugate Gaussian model on the $\log{(HR)}$ appears to be a reliable and straightforward method to obtain Bayesian estimates for survival data, provided a proportional hazard setting. In such situations, the main usefulness of the Bayesian approaches may be enabling probabilistic statements that could not be obtained otherwise. Therefore, Bayesian approaches should be used more often.

\section*{Computational details}

The results in this paper were obtained using R~3.5.3 (distributed under the terms of the GNU General public License Version 2 or 3), with the following packages: 
\texttt{survival}~2.43-3, \texttt{spBayesSurv}~1.1.3, \texttt{coda}~0.19.2, \texttt{bayesplot}~1.7.0, \texttt{INLA}~18.07.12, \texttt{survHE}~1.0.65, \texttt{SemiCompRisks}~3.3, \texttt{dynsurv}~0.3-6, \texttt{rstantools}~1.5.1, \texttt{rstan}~2.18.2, \texttt{rjags}~4-8. R itself and packages used are available from the Comprehensive R Archive Network (CRAN) at \emph{https://CRAN.R-project.org/}, except for the \texttt{INLA} package, which can be installed directly from the R-\texttt{INLA} project website (\emph{install.packages("INLA", repos=c(getOption("repos"), INLA="https://inla.r-inla-download.org/R/stable"), dep=TRUE)}) and the \texttt{rstanarm} package, whose development version including the \emph{stan$\_$surv()} function used for the reported analyses, can be downloaded and installed from Github (As of 22 July 2019, \emph{devtools}{::}\emph{install$\_$github("stan-dev/rstanarm", ref = "feature/survival", build $\_$vignettes = FALSE)}).

\section*{Additional information}
The authors declare no potential conflicts of interest with respect to the research, authorship, and publication of this article.

\section*{Funding}
LB was funded by the TRT$\_$cSVD project (\emph{From Target Identification to Next Generation Therapies for Cerebral Small Vessel Diseases}, Pr Hugues Chabriat \& Dr Anne Joutel, \emph{RHU - Agence Nationale pour la Recherche}, grant number: ANR-16-RHUS-004).

\bibliographystyle{vancouver}
\bibliography{biblio}

\clearpage
\section{Supplementary data}
\subsection{Default MCMC parameterizations}

We initially applied the available R packages and functions using their respective default parameterizations. 

Table \ref{tab:mcmcparameters_default} reports the default parameters for Markov chain Monte Carlo (MCMC) simulations  for the functions (total number of iterations, number of iterations for the burn-in sequence, thinning value, number of chains).

The approach of Spiegelhalter \emph{et al.} (1994) and INLA estimation rely on other methods for estimation (conjugate Gaussian prior and integrated nested Laplace approximation, respectively) and are therefore not included in this table.

\begin{table}[!h]
    \caption{Monte Carlo simulation default parameterizations in the R packages for Bayesian survival PH models}
    \label{tab:mcmcparameters_default}
\begin{footnotesize}
\centering
    \begin{tabular}{lccccl}
    \hline
    &\multicolumn{2}{c}{Number of iterations} &&& \\ \cline{2-3}
      Package/Method  & Total& Burn-in & Thin & No. of chains & Comment\\
        \hline
        \multicolumn{6}{l}{\underline{\texttt{SemiCompRisks} package}}\\
        \emph{BayesSurv$\_$HReg()}& 2000 & 1000 & 10 & 2& Model output reports  posterior \\
       &&&&&median estimates (not mean)\\
       &&&&&\\ 
  \multicolumn{6}{l}{\underline{\texttt{survHE} package}}\\      
\emph{fit.models(...,method="hmc")}& 2000 & 1000 & 1  &2& \\
  \multicolumn{6}{l}{\underline{\texttt{dynsurv} package}}\\
          \emph{bayesCox()} & 3000 & 500 & 1 & 1& MCMC output includes all \\
       &&&&&iterations (burn-in included\\
       &&&&&and before thinning)   \\
&&&&&\\ 
 \multicolumn{6}{l}{\underline{\texttt{rstanarm} package$^\star$}}\\
					\emph{stan$\_$surv()}	&  2000 & 1000  & 1 &  4 &  \\

&&&&&\\ 
\multicolumn{6}{l}{\underline{\texttt{spBayesSurv} package}}\\
        \emph{indeptCoxph()}&  5000 & 3000  & 1 &  1  &  \\
         \emph{survregbayes()}& 5000 & 3000  & 1 &  1  & \\
         \emph{survregbayes2()}&5000 & 3000  & 1 &  1  & \\
&&&&&\\ 
\multicolumn{6}{l}{\underline{Murray et al (2016)}}\\
                                & 12000 & 2000& 1& 2&\\ 
 
         \hline
    \end{tabular}

\end{footnotesize}
\end{table}
{\footnotesize $^\star$ As of 22 July 2019, \texttt{rstanarm} development version, including the \emph{stan$\_$surv()} function, downloaded from the survival branch of the package available on github at https://github.com/stan-dev/rstanarm/tree/feature/survival}

\subsection{Numerical example}
\subsubsection{Estimates with default parameterizations}
Table \ref{tab:loghr2} presents the estimates of the log hazard ratio (HR) of event (death or airflow decline) in the azithromycin group over placebo according to the method on data derived from the ALLOZITHRO trial dataset, obtained with the different methods: conventional frequentist methods and the Bayesian survival PH models. Functions for Bayesian inference were first applied with their respective default parameterization, reported in table \ref{tab:loghr2}. 

\begin{figure}[!h]
\centering
    \includegraphics[width=4.5in]{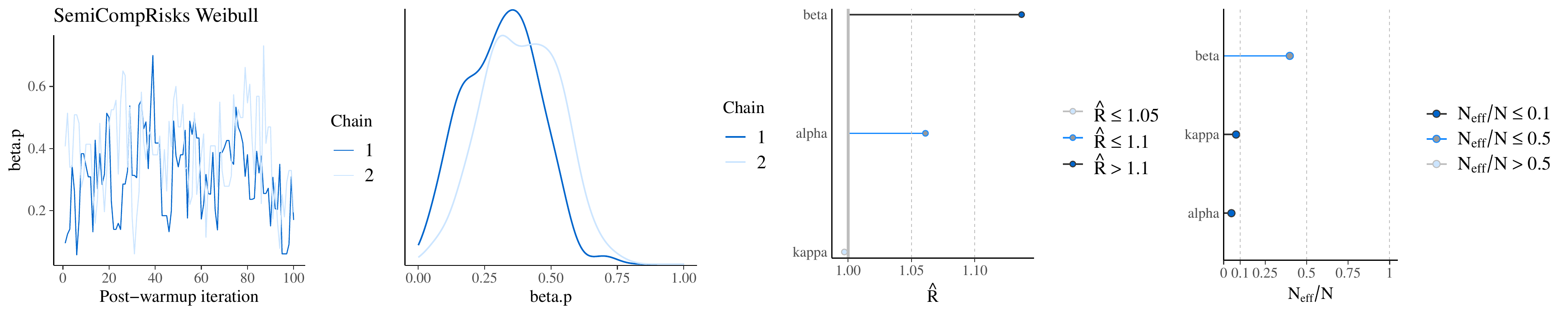}
    \includegraphics[width=4.5in]{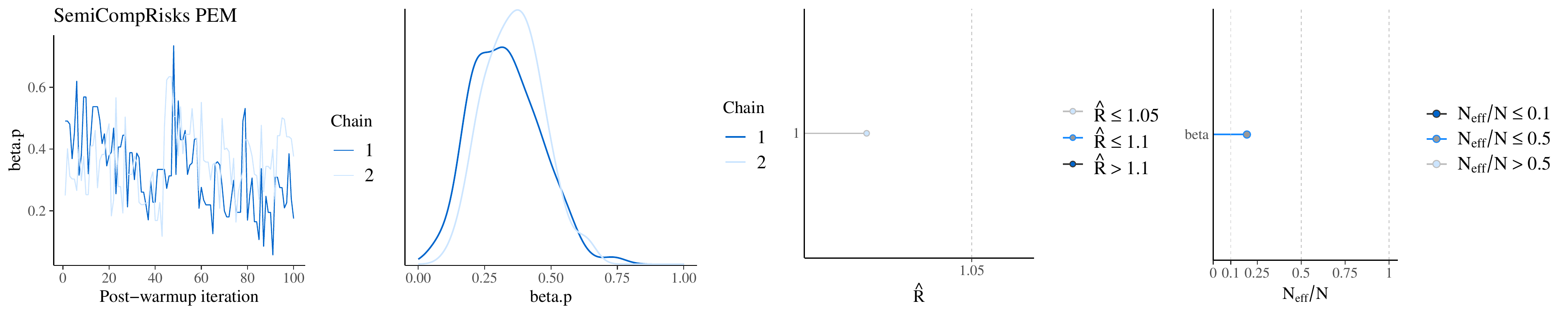}
    \includegraphics[width=4.5in]{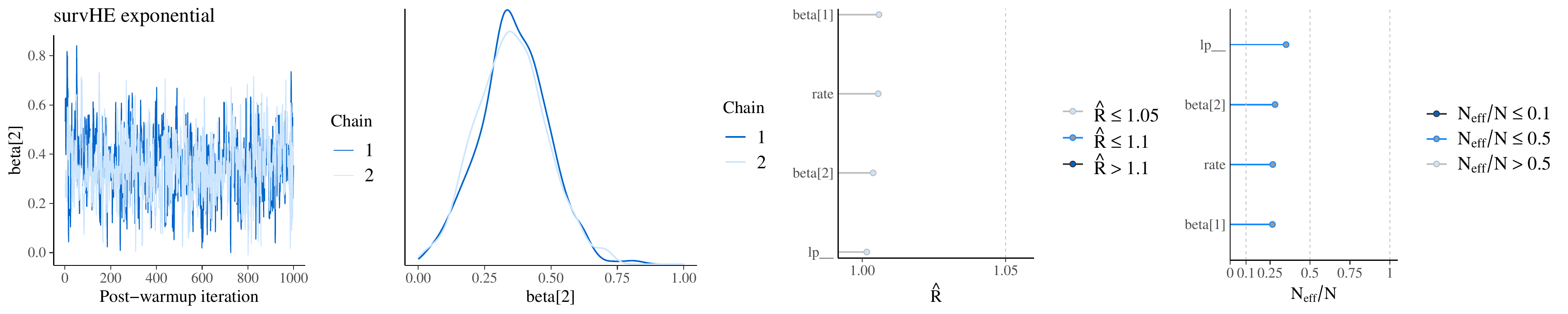}
    \includegraphics[width=4.5in]{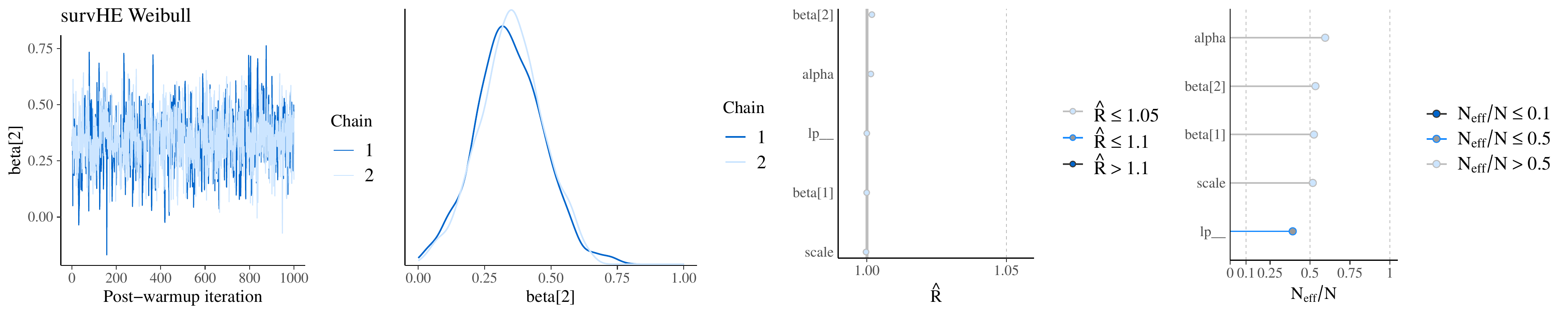}  
    \includegraphics[width=4.5in]{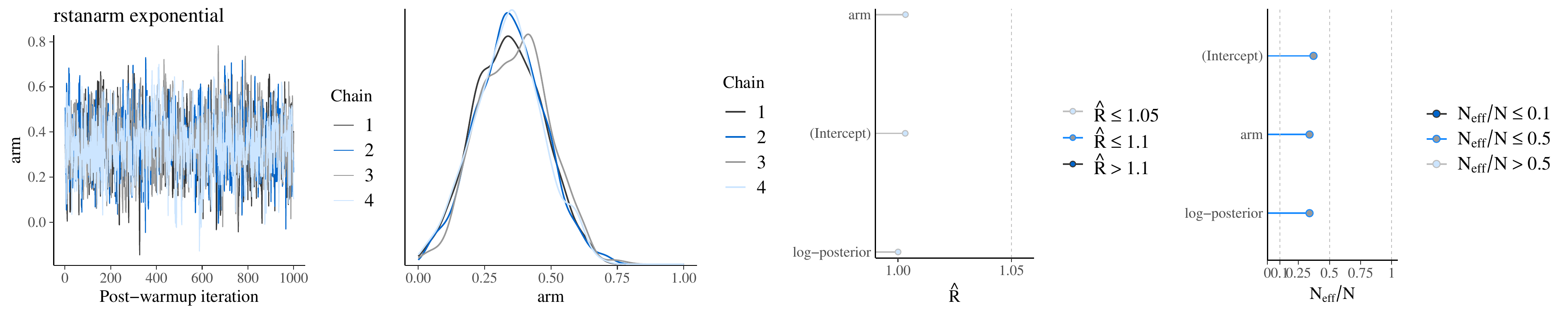}
    \includegraphics[width=4.5in]{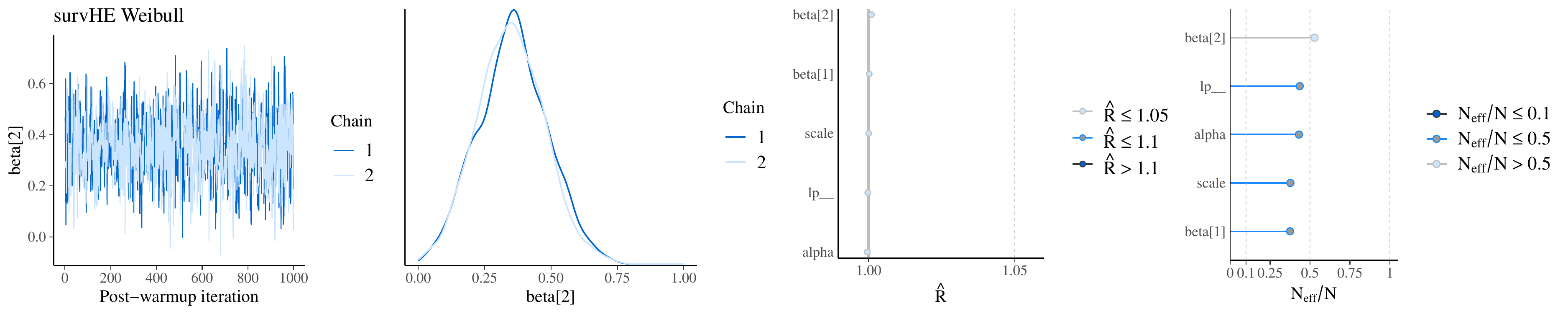}  
		\includegraphics[width=4.5in]{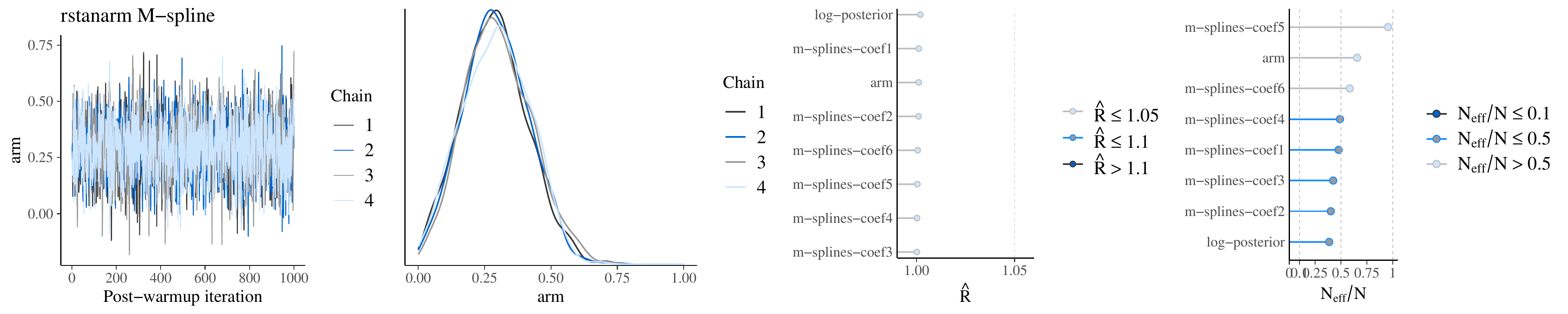} 
		\includegraphics[width=4.5in]{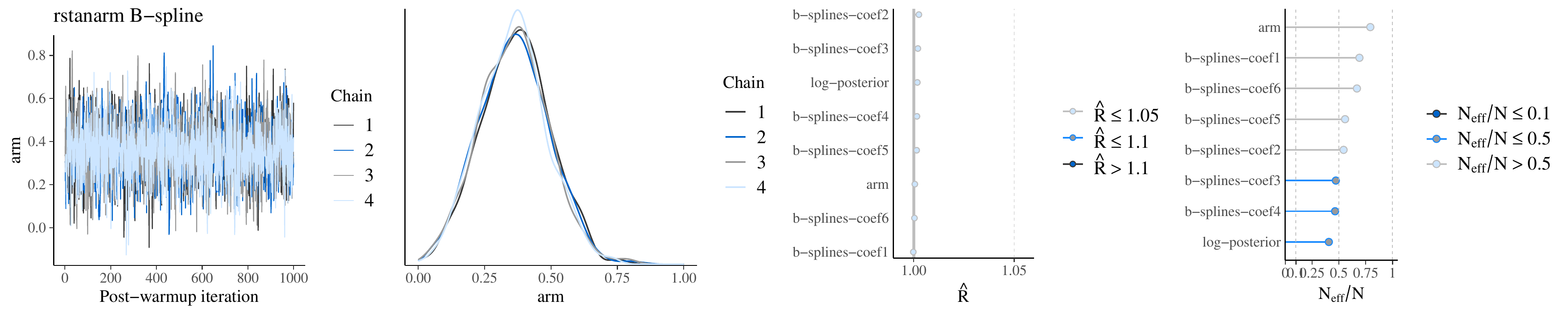} 
    \includegraphics[width=4.5in]{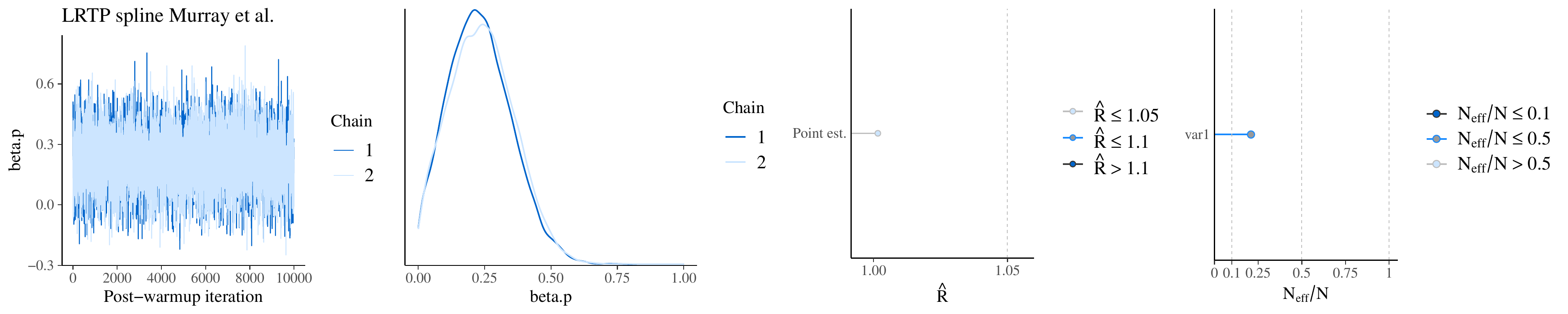}  
    \caption{Convergence diagnostics in MCMC with default parameterization (1/2): from top to bottom, \texttt{SemiCompRisks} package's functions \emph{BayesSurv$\_$HReg(..., model="Weibull"), BayesSurv$\_$HReg(..., model='PEM')}, \texttt{survHE} package's functions \emph{fit.models(..., method="hmc", distr="exponential"), fit.models(..., method="hmc", distr="weibullPH")}, \texttt{rstanarm} package's functions \emph{stan$\_$surv(..., basehaz="exponential"), stan$\_$surv(..., basehaz="weibull"), stan$\_$surv(..., basehaz="ms"), stan$\_$surv(..., basehaz="bs")}, Murray et al (2016)'s model.} \label{fig-diag_default1}
\end{figure}

\begin{figure}[ht!]
\centering
		\includegraphics[width=3.75in]{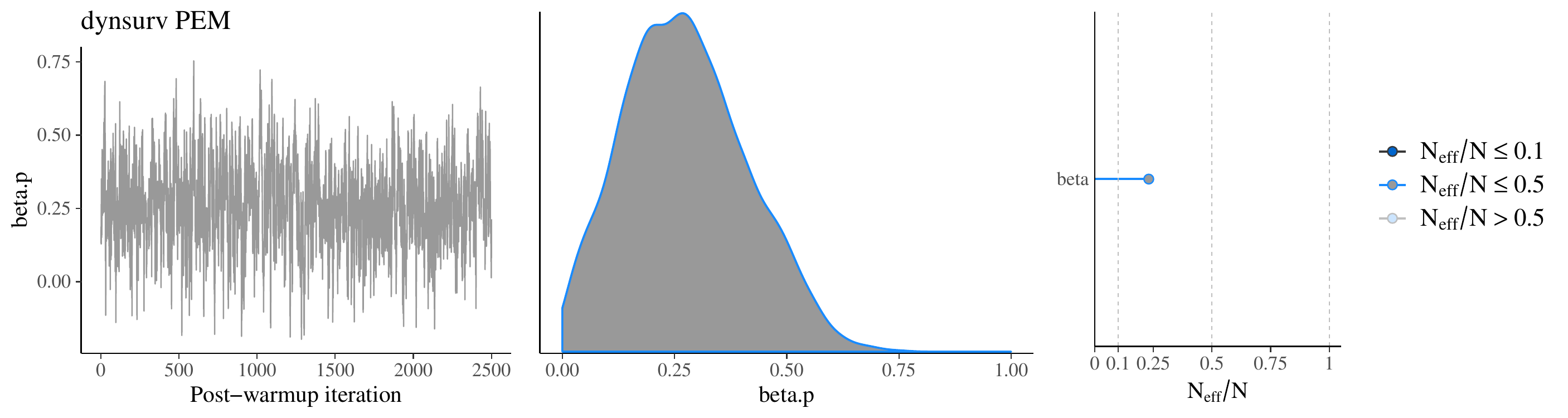} 
		\includegraphics[width=3.75in]{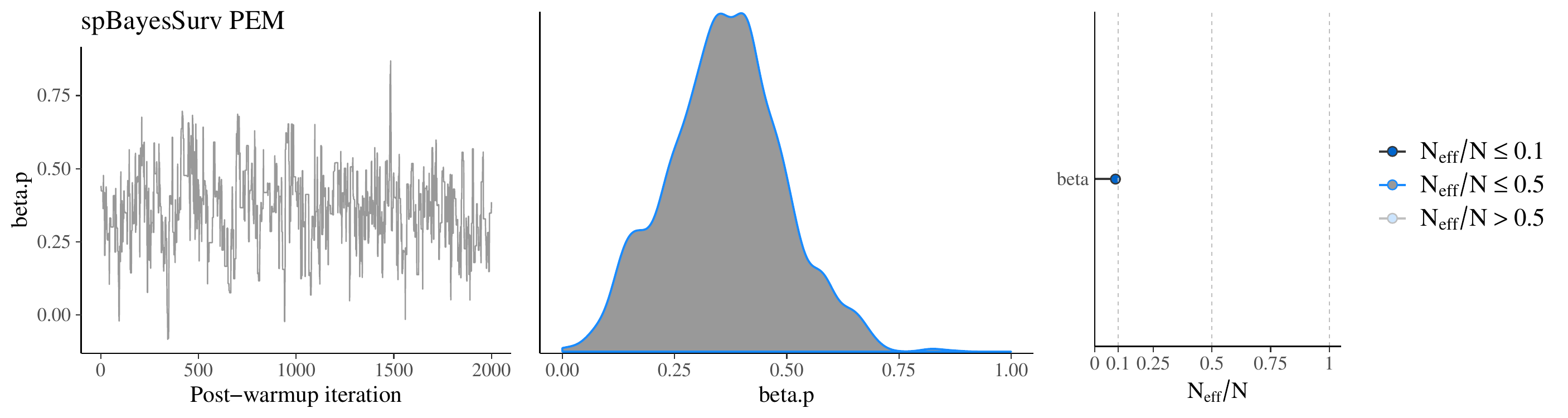}
    \includegraphics[width=3.75in]{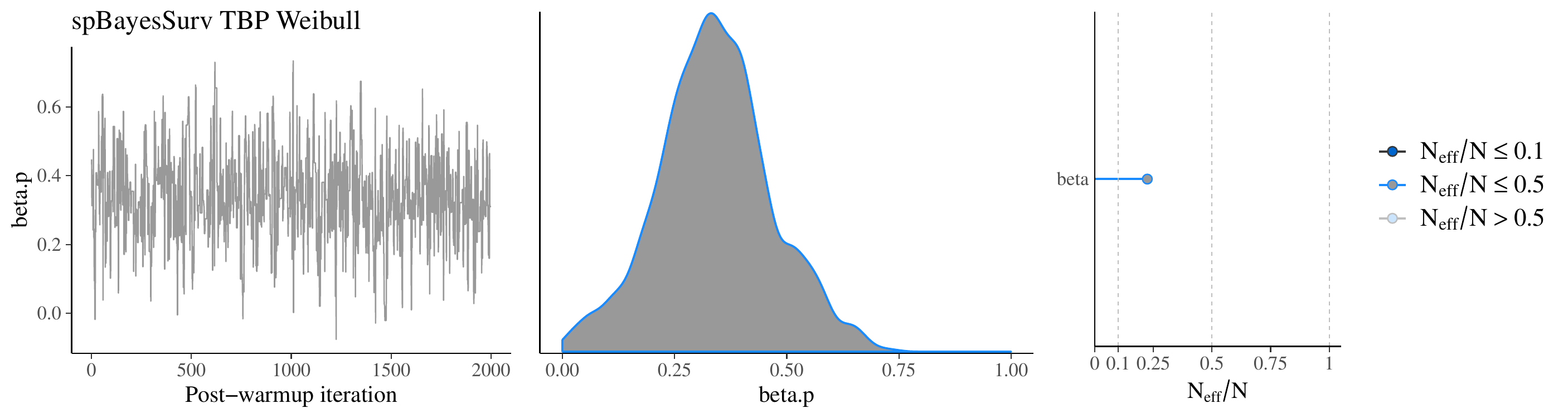}
    \includegraphics[width=3.75in]{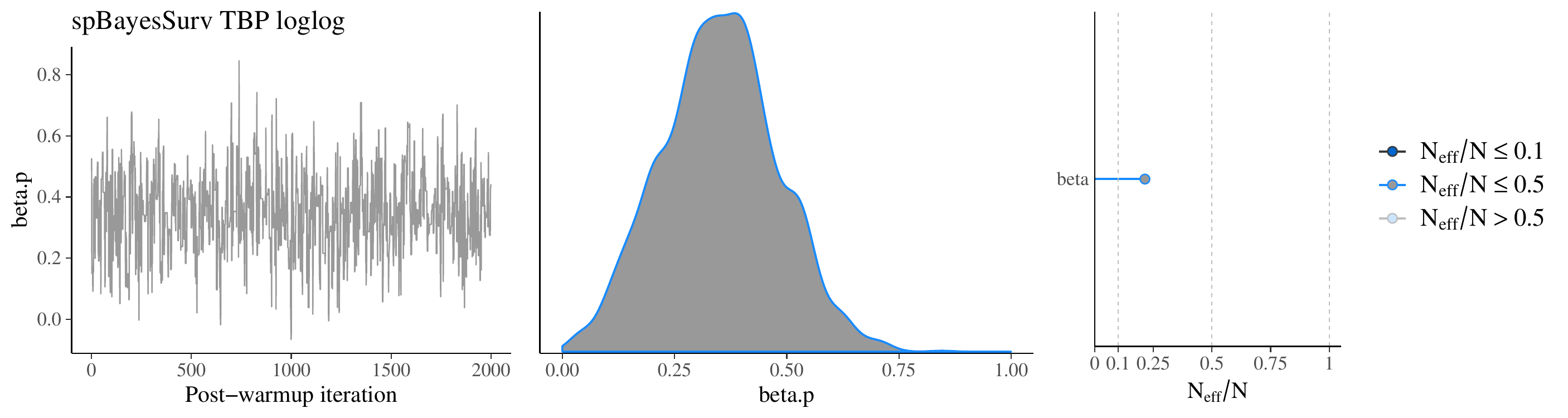}
    \includegraphics[width=3.75in]{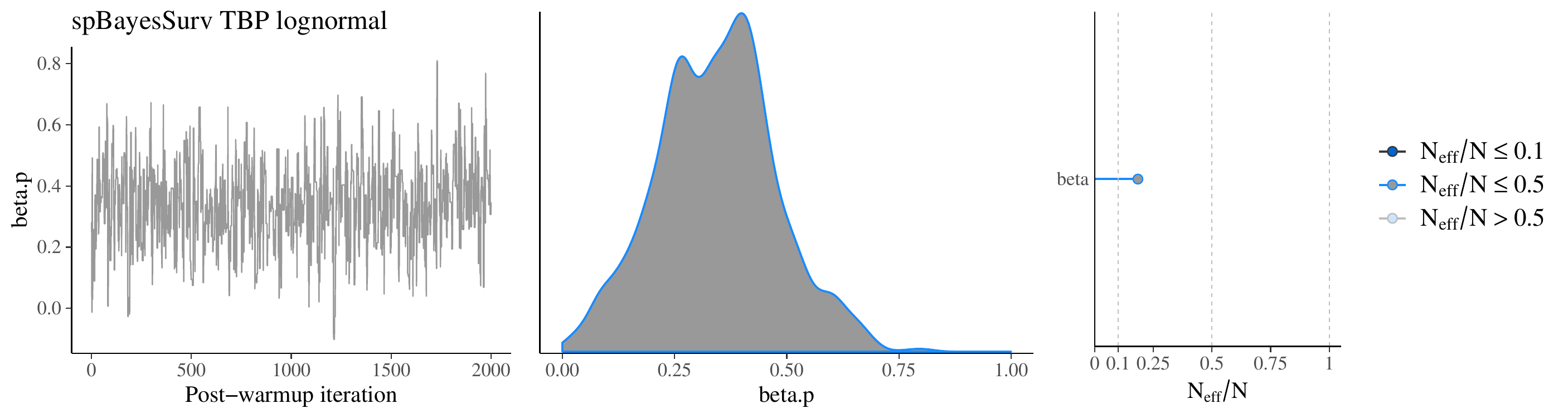}
    \includegraphics[width=3.75in]{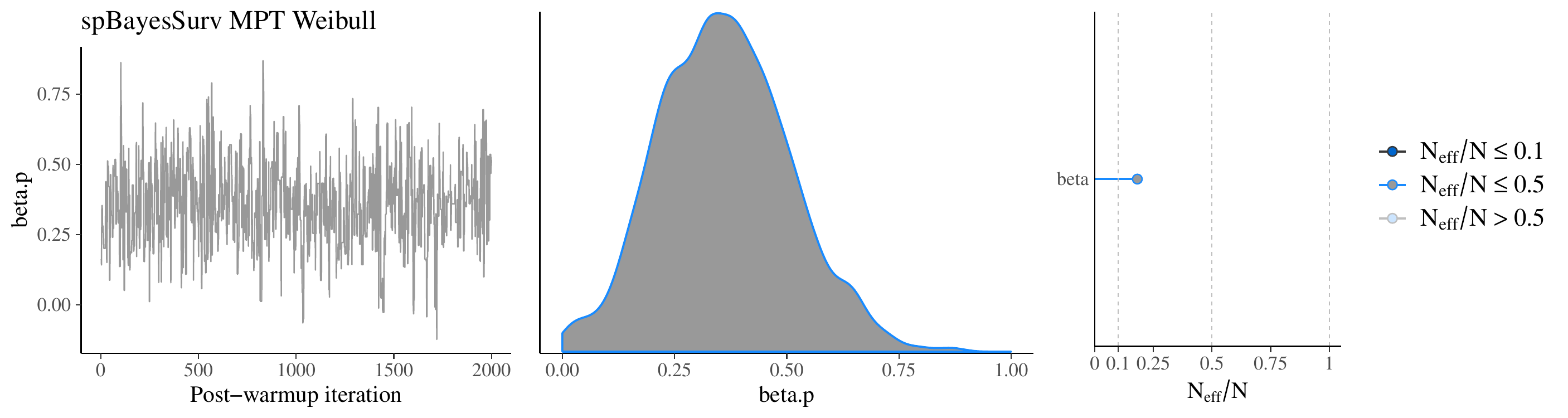}
    \includegraphics[width=3.75in]{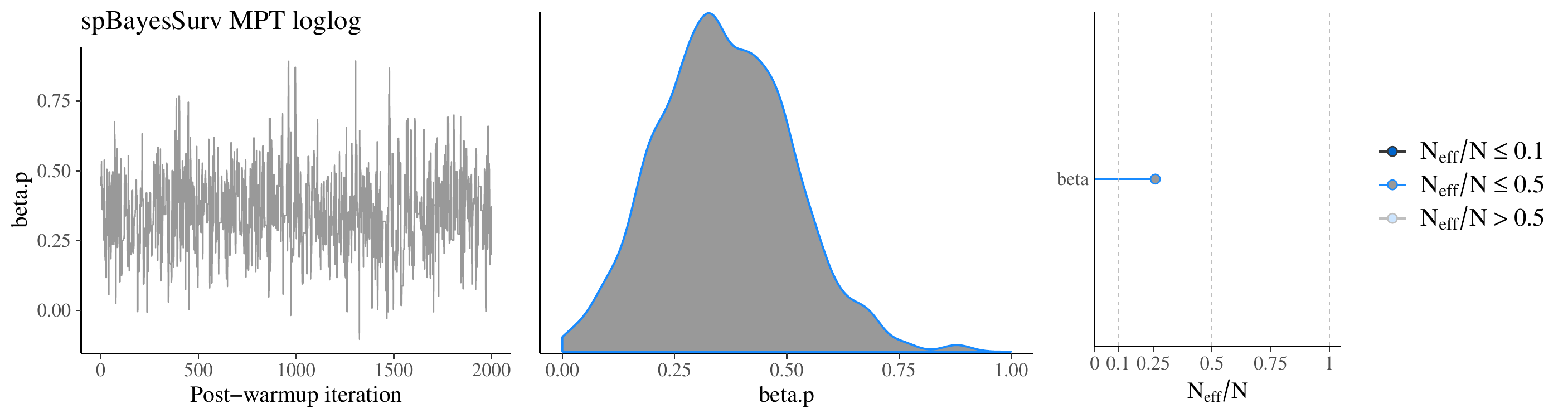}
    \includegraphics[width=3.75in]{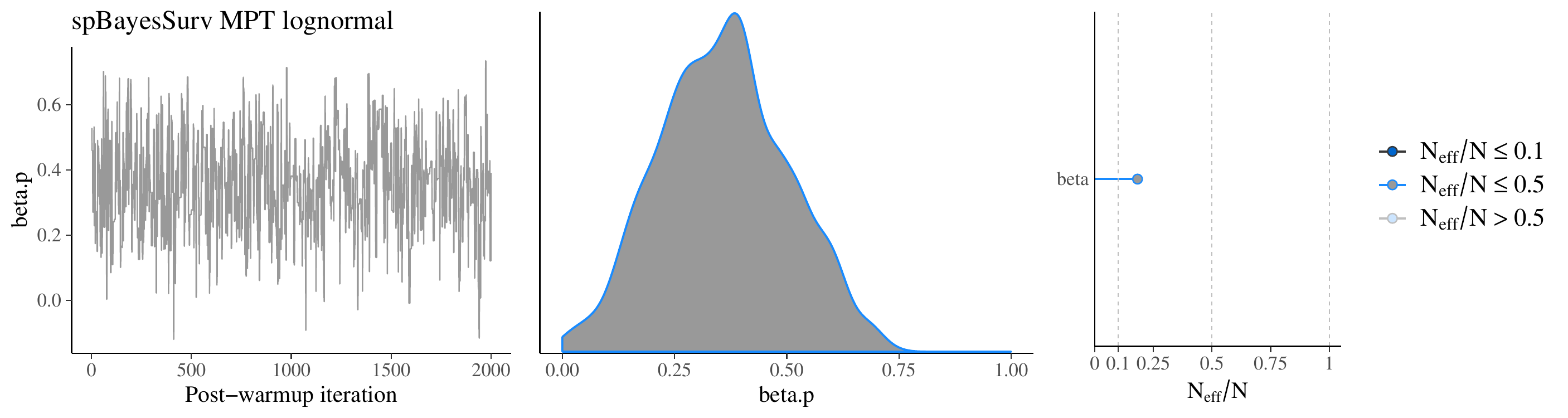}
    \caption{Convergence diagnostics in MCMC with default parameterization (2/2): \texttt{dynsurv} package function \emph{bayesCox(..., model = "TimeIndep")}, \texttt{spBayesSurv} package, from top to bottom, functions \emph{indeptCoxph(), survreg(..., survmodel="PH", dist="weibull"), survreg(..., survmodel="PH", dist="loglogistic"), survreg(..., survmodel='PH', dist="lognormal"), survreg2(..., survmodel="PH", dist="weibull"), survreg2(..., survmodel="PH", dist="loglogistic"), survreg(..., survmodel="PH", dist="lognormal")}. } \label{fig-diag_default2}
\end{figure}

Figure \ref{fig-diag_default1} and \ref{fig-diag_default2} illustrate basic convergence diagnostics on models fitted using MCMC simulations, with default paramterizations of the respective functions. Each line of the figures corresponds to one function: from left to right, traceplot of $log(HR)$ posterior draw by iteration per chain, estimated posterior density of $log(HR)$ per chain, potential scale reduction factor (PSRF) or ${\hat R}$ across chains (computed when multiple MCMC chains), effective sample size to number of saved iterations ratio ($N_{eff}/N$) over all chains. Diagnostics and plots were obtained using \texttt{coda} and \texttt{bayesplot} packages. For homogeneity purposes, we applied the same diagnostics tools across all functions, and in particular $\hat{R}$ and $N_{eff}/N$ ratio were computed using the functions \emph{effectiveSize()} and \emph{gelman.diag()} in \texttt{coda}, although specific estimates are directly available in the output of models fitted with HMC/Stan (obtained here via the \texttt{survHE} package). Similarly, specific tools are available for models estimated via HMC simulations (Stan).

Functions in the \texttt{SpBayesSurv} package runs only one MCMC chain by default; ${\hat R}$ could not be computed (Figure \ref{fig-diag_default2}).

\begin{table}[!h]
\begin{small}
    \centering
        \caption{ALLOZITHRO Trial: Estimates of the log hazard ratio (HR) of event (death or airflow decline) in the azithromycin group over placebo according to the method. 95\%CI refers to the confidence interval for frequentist estimates and to credible intervals for the Bayesian models. Bayesian models were estimated with the default parameterization of the respective functions. The probability of a hazard ratio greater than 1.5 can be computed only using Bayesian analyses. 
  PEM: piecewise exponential model; RW: random walk; TBP: transformed Bernstein polynomial; MPT: mixture of Polya trees}\label{tab:loghr2}
    \begin{tabular}{lccc}
    \hline
         Estimation method  & \multicolumn{2}{c}{Posterior $log HR$}& \\ \cline{2-3}
       & Mean& 95\% CI& $Pr (HR>1.5)$\\
        \hline
         \multicolumn{3}{l}{\textbf{Frequentist approach}}\\
        Semiparametric Cox &0.3661 & 0.1056 to 0.6265&-\\
        Parametric exponential & 0.3493 & 0.0896 to 0.6090  &-\\
        Parametric Weibull & 0.3389 & 0.0886 to 0.5893 &-\\
         &  &  &  \\
        \multicolumn{3}{l}{\textbf{Bayesian models}}\\
        \multicolumn{3}{l}{\underline{Spiegelhalter et al. (1994)}}\\
        \quad Skeptical prior &   0.3506 &  0.0957 to 0.6055 & 0.3366\\
        \quad Enthusiastic prior & 0.3317 & 0.0769 to 0.5866 & 0.2854\\
        
				\multicolumn{3}{l}{\underline{\texttt{SemiCompRisks} package}}\\
        Parametric Weibull &  0.3507 & 0.0773 to 0.6082 & 0.3750 \\
        PEM &  0.3454 & 0.1623 to 0.5700 & 0.3100 \\
				
        \multicolumn{3}{l}{\underline{\texttt{INLA} package}}\\
        Parametric exponential &  0.3491 & 0.0900 to 0.6102 & 0.3346\\
        Parametric Weibull & 0.3531 & 0.0937 to 0.6144 & 0.3456\\        
        PEM RW1 & 0.3479 & 0.0883 to 0.6094 & 0.3317\\      
				PEM RW2 & 0.3481 & 0.0885 to 0.6096 & 0.3322\\      
        
		\multicolumn{3}{l}{\underline{Stan via \texttt{survHE} package}}\\
        Parametric exponential & 0.3549 & 0.0968 to 0.6148 & 0.3465\\
        Parametric Weibull &  0.3510 & 0.0992 to 0.6070 & 0.3315\\        

		\multicolumn{3}{l}{\underline{Stan via \texttt{rstanarm} package}$^\star$}\\
        Parametric exponential & 0.3446 & 0.0810 to 0.6032 & 0.3275\\
        Parametric Weibull  & 0.3599 & 0.1104 to 0.6065 &0.3525\\        
				M-splines  & 0.2890 & 0.0405 to 0.5366 & 0.1905\\
        B-splines  & 0.3599 & 0.1041 to 0.6175 & 0.3605\\    
				
          \multicolumn{3}{l}{\underline{\texttt{dynsurv} package}}\\
           PEM Gamma & 0.2604 & -0.0415 to 0.5531 & 0.1720\\
         
		\multicolumn{3}{l}{\underline{\texttt{spBayesSurv} package}}\\
        PEM \emph{indeptCoxph()}& 0.3608 & 0.1163 to 0.6300  & 0.3660\\
    
			TBP \emph{survregbayes()}& &\\
       \quad Weibull     & 0.3373 & 0.0618 to 0.5903& 0.2760\\
       \quad log-logistic & 0.3490 &0.0991 to 0.6088 & 0.3350\\
       \quad log-normal  & 0.3413 &0.0748 to 0.6195 & 0.3195\\
     
		MPT  \emph{survregbayes2()} & &  \\
        \quad Weibull & 0.3558 & 0.0515 to 0.6524 & 0.3580\\
         \quad log-logistic & 0.3584 & 0.0743 to 0.6713 & 0.3790\\
        \quad log-normal  & 0.3555 & 0.0665 to 0.6292 & 0.3535\\
    
		\multicolumn{3}{l}{\underline{Murray \emph{et al.} (2016)}}\\
   LRTP Splines & 0.2233 & -0.0352 to 0.4833 & 0.0841\\
\hline
    \end{tabular}

\end{small}
{\footnotesize $^\star$ As of 22 July 2019, \texttt{rstanarm} development version, including the \emph{stan$\_$surv()} function, downloaded from the survival branch of the package available on github at https://github.com/stan-dev/rstanarm/tree/feature/survival}
\end{table}

\clearpage
\subsubsection{Sensitivity analysis with optimized parameterizations}
A sensitivity analysis was conducted with optimized MCMC parameterization to ensure convergence for the estimation of posterior distributions (for functions using MCMC estimation).

\begin{figure}[!h]
\centering
    \includegraphics[width=4.5in]{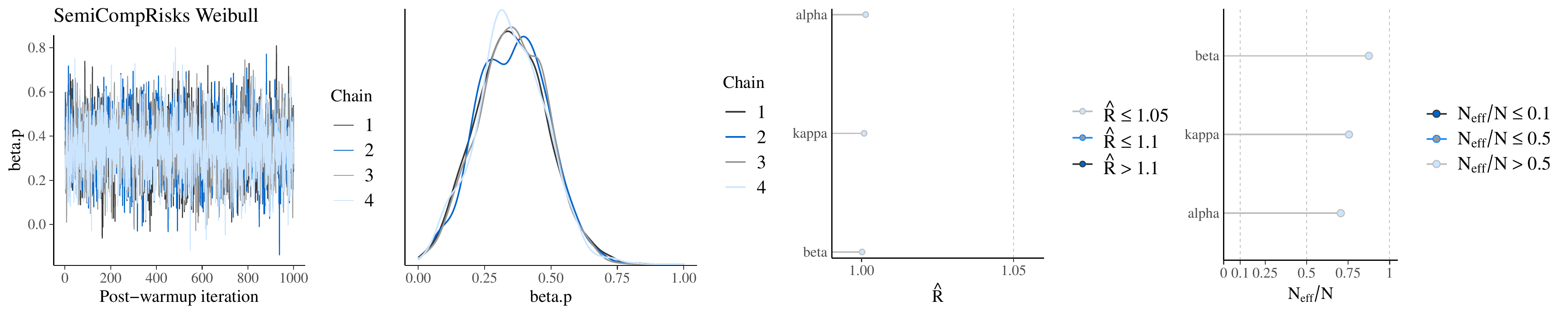}
    \includegraphics[width=4.5in]{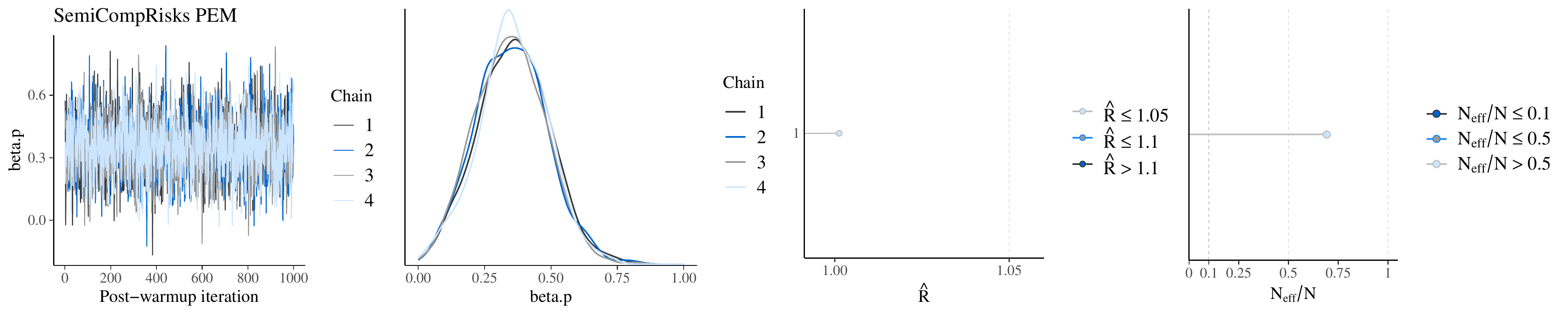}
    \includegraphics[width=4.5in]{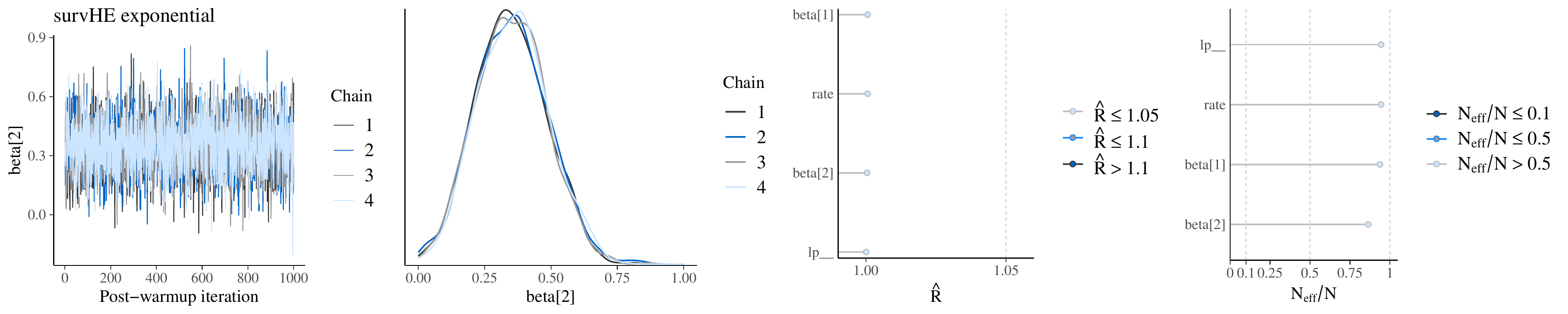}
    \includegraphics[width=4.5in]{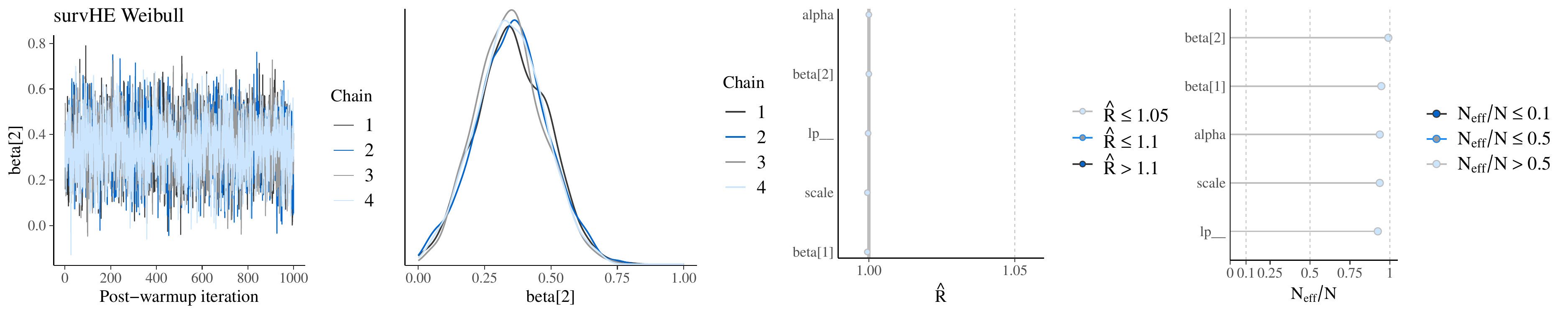}  
		\includegraphics[width=4.5in]{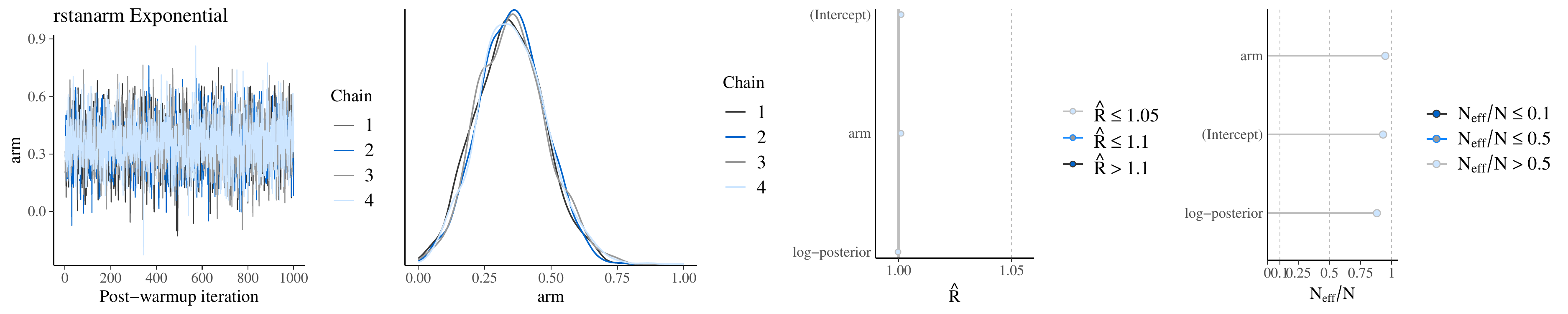} 
		\includegraphics[width=4.5in]{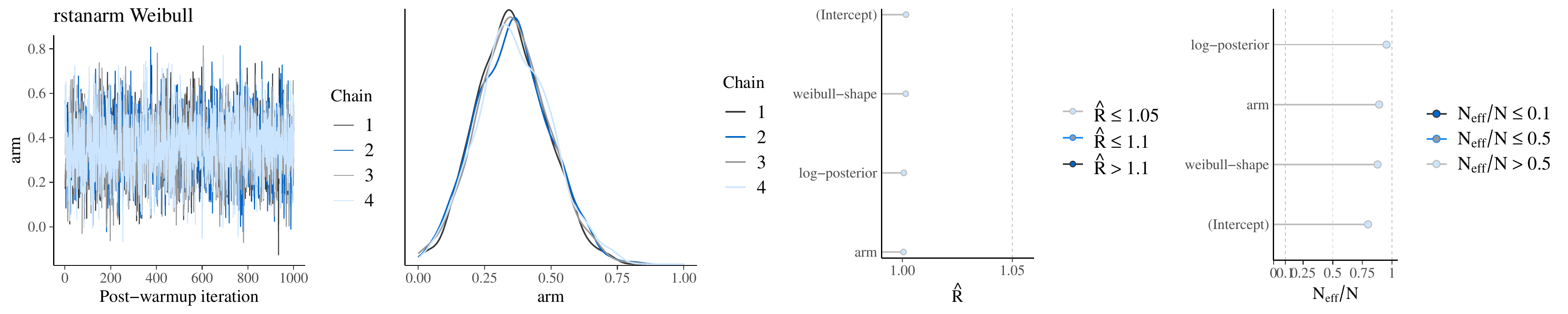} 
		\includegraphics[width=4.5in]{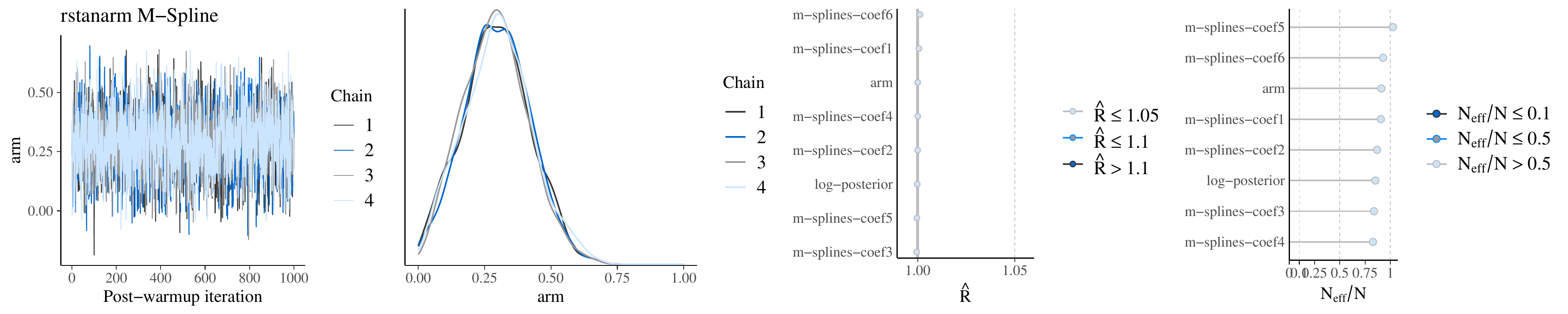} 
		\includegraphics[width=4.5in]{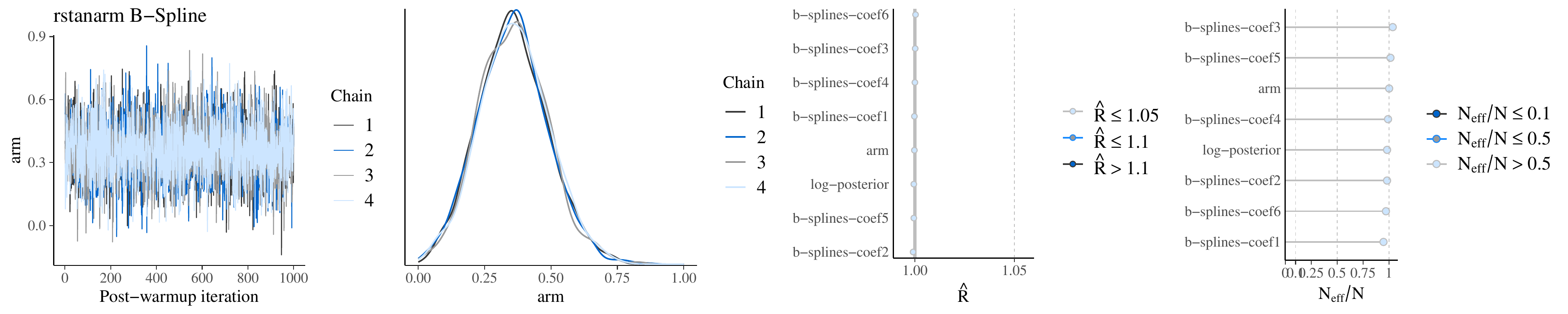} 
    \includegraphics[width=4.5in]{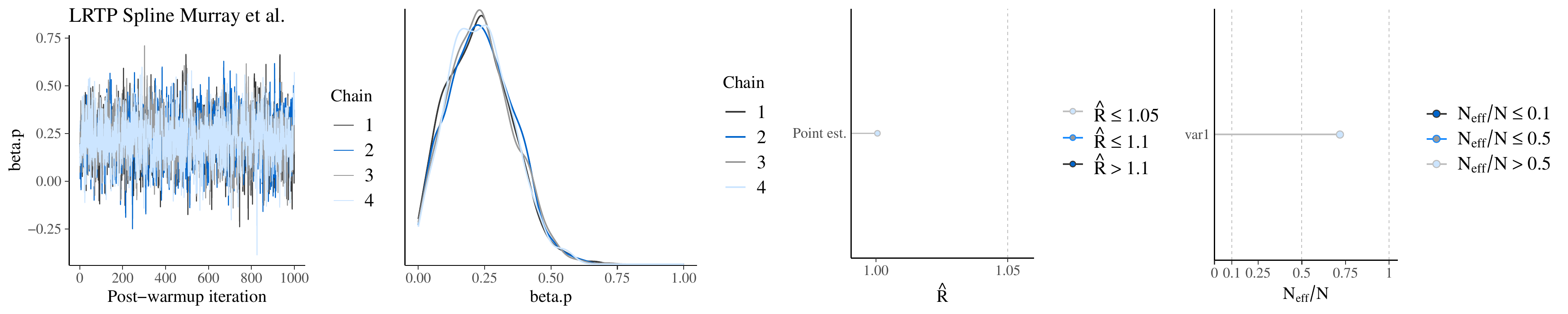}
    \caption{Convergence diagnostics in sensitivity analysis for optimized MCMC: from top to bottom, \texttt{SemiCompRisks} package's functions \emph{BayesSurv$\_$HReg(..., model="Weibull"), BayesSurv$\_$HReg(..., model="PEM")}, \texttt{survHE} package's functions \emph{fit.models(..., method="hmc", distr="exponential"), fit.models(..., method="hmc", distr="weibullPH")}, \texttt{dynsurv} package function \emph{bayesCox(..., model = "TimeIndep")}, Murray et al (2016)'s model .} \label{fig-diag_opti1}
\end{figure}

\begin{figure}[ht!]
\centering
		\includegraphics[width=4.5in]{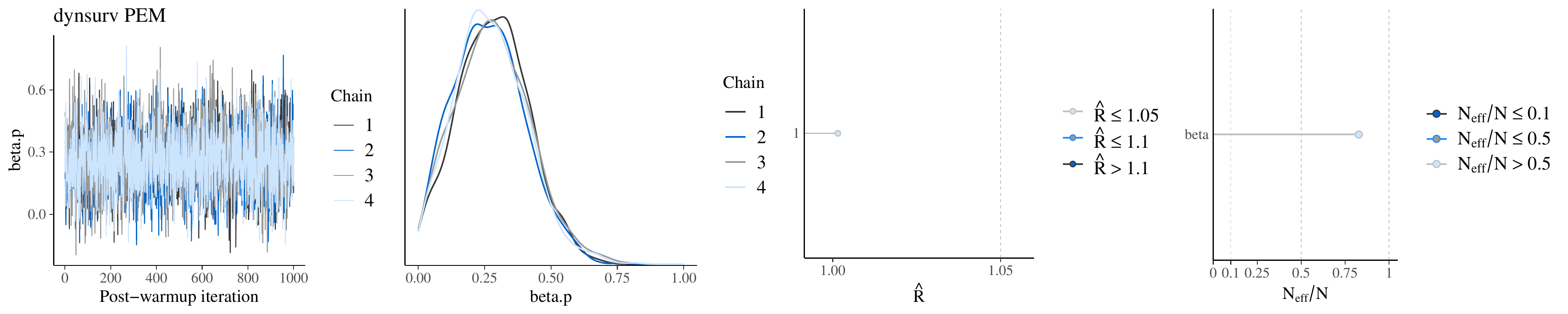}
		\includegraphics[width=4.5in]{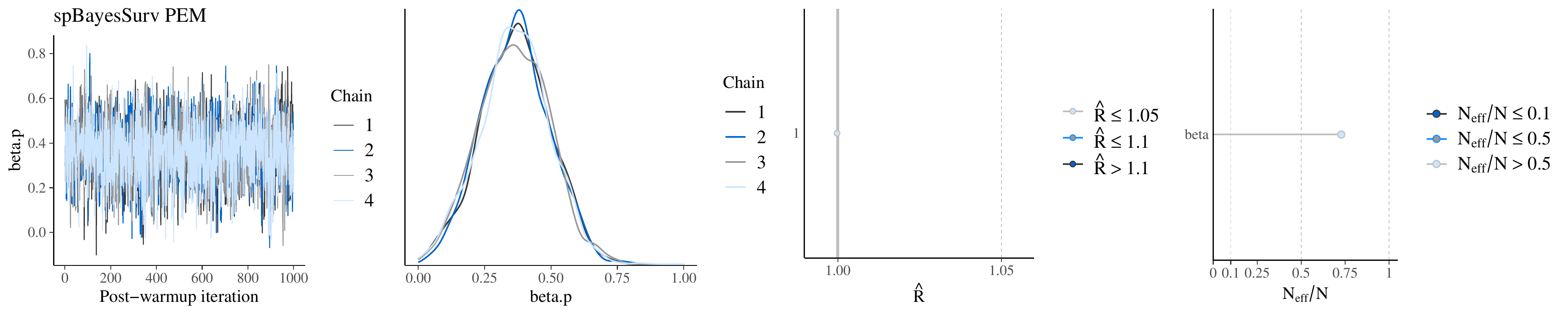}
    \includegraphics[width=4.5in]{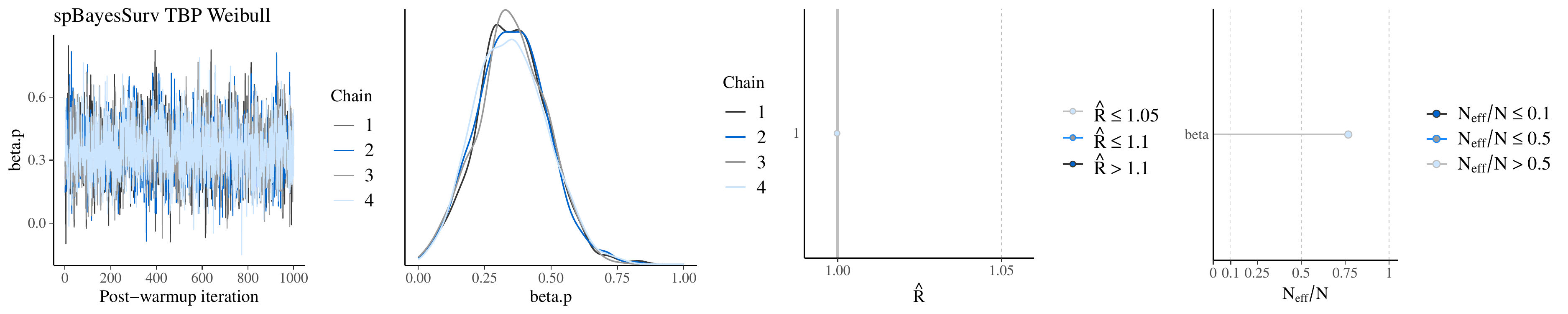}
    \includegraphics[width=4.5in]{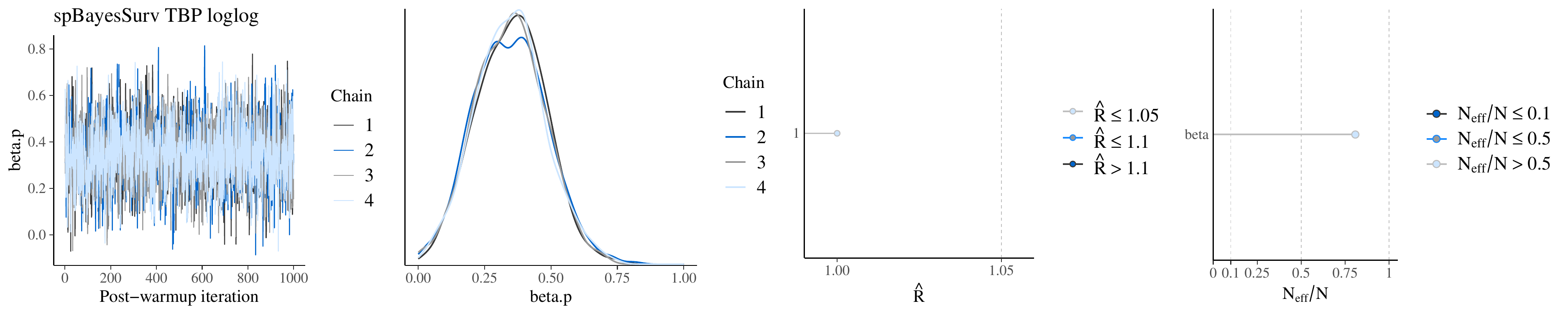}
    \includegraphics[width=4.5in]{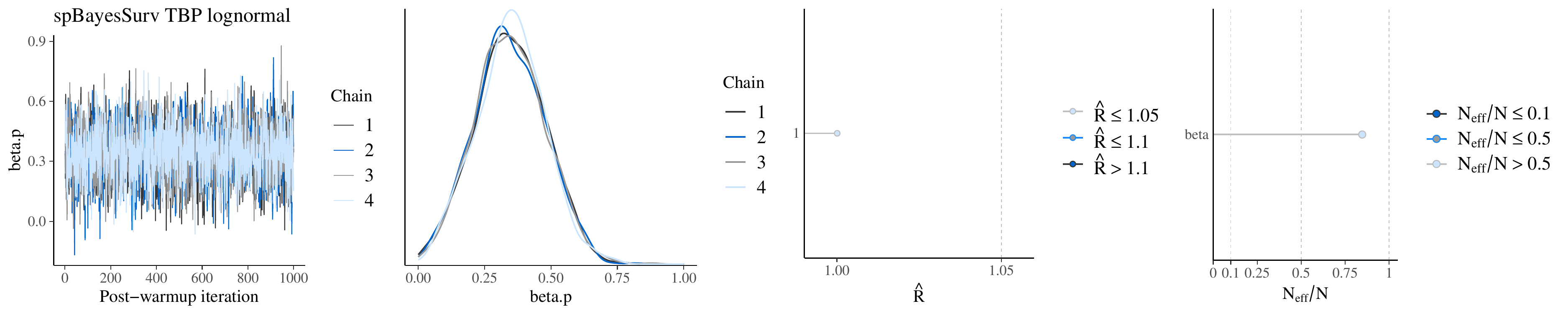}
    \includegraphics[width=4.5in]{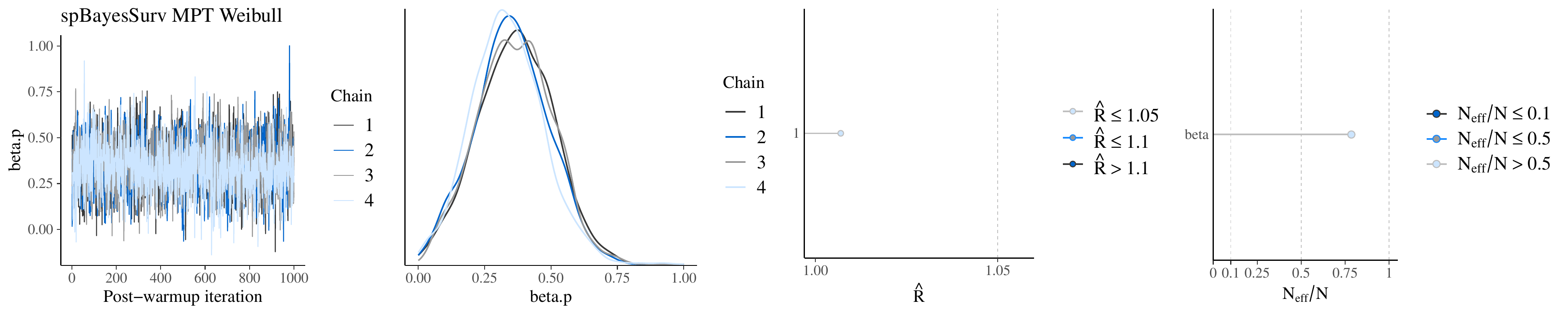}
    \includegraphics[width=4.5in]{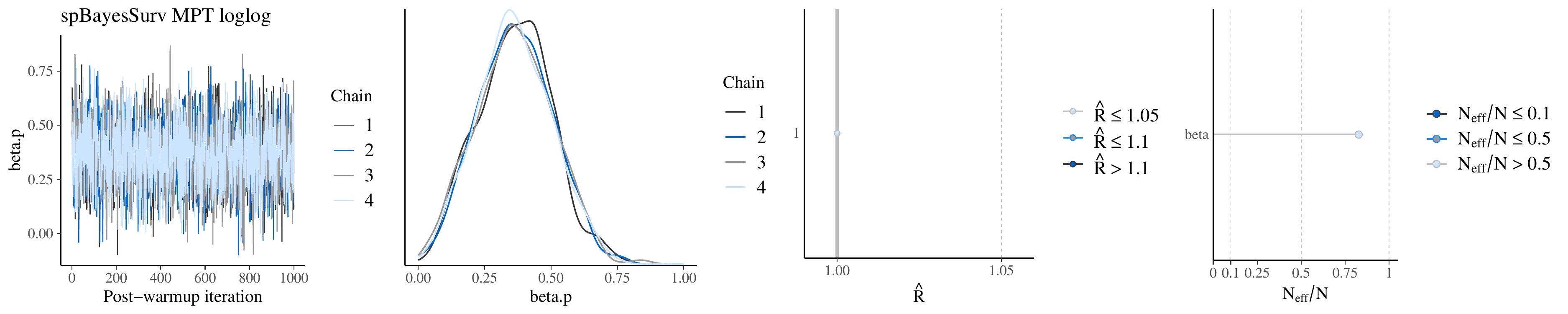}
    \includegraphics[width=4.5in]{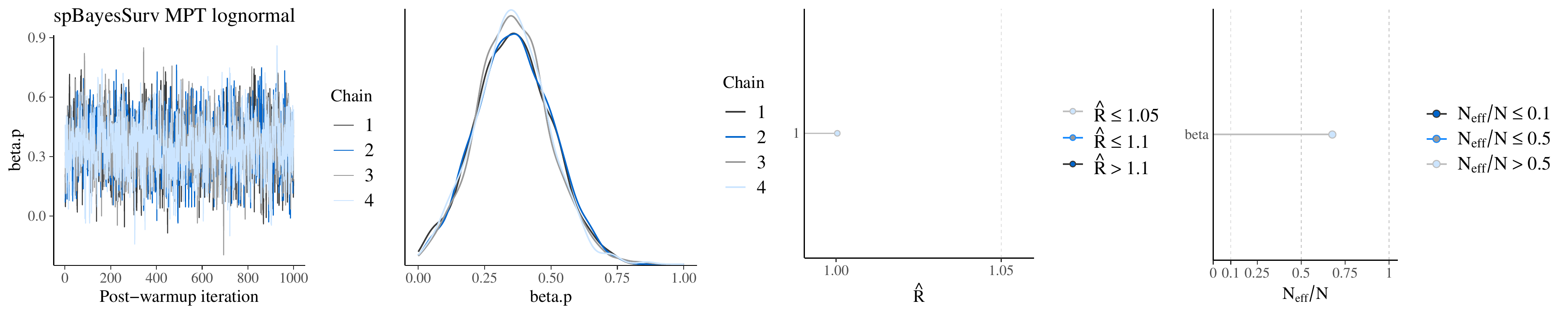}
    \caption{Convergence diagnostics in sensitivity analysis for optimized MCMC: \texttt{spBayesSurv} package, from top to bottom, functions \emph{indeptCoxph(), survreg(..., survmodel="PH", dist="weibull"), survreg(..., survmodel="PH", dist="loglogistic"), survreg(..., survmodel="PH", dist="lognormal"), survreg2(..., survmodel="PH", dist="weibull"), survreg2(..., survmodel="PH", dist="loglogistic"), survreg(..., survmodel="PH", dist="lognormal")}. } \label{fig-diag_opti2}
\end{figure}

We used 4 chains and set the number of iterations per chain and thinning interval in order to fulfill the following conditions on MCMC convergence: burn-in representing half of the iterations, at least 1000 saved iterations per chain, effective sample size to total number of iterations ratio ($N_{eff}:N$ ratio) greater than 0.5. Table \ref{tab:mcmcparameters} reports the optimized parametrizations which were necessary to fullfill these criteria (total number of iterations per chain, including burn-in and before thinning, and thinning interval). Figures \ref{fig-diag_opti1} and \ref{fig-diag_opti2} illustrate convergence for the different functions obtained with these optimized MCMC parameters.

\begin{figure}[!h]
\centering
    \includegraphics[width=6in]{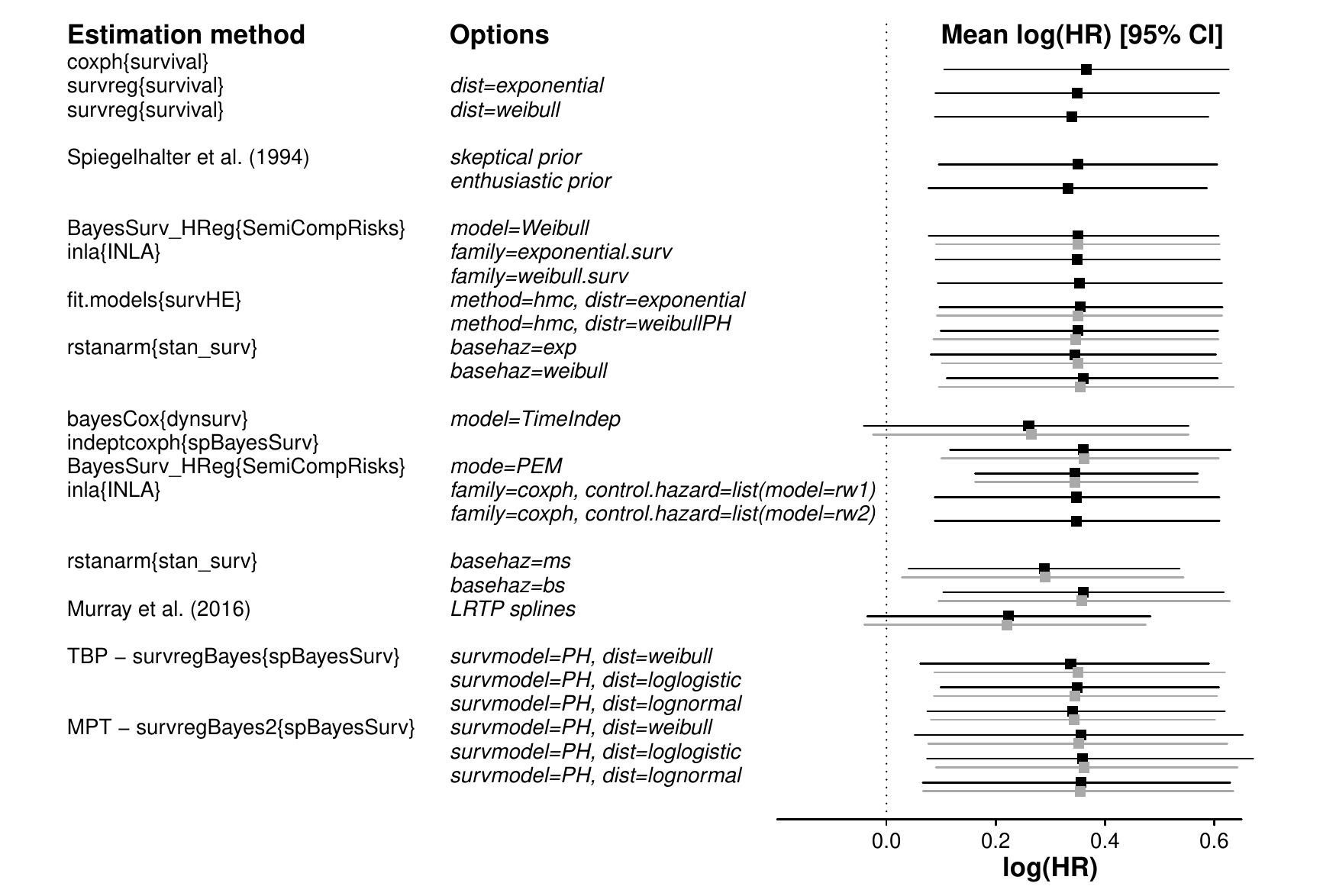}
   
    \caption{ALLOZITHRO Trial: Forest plot of $\log{(HR)}$ posterior estimates by type of model and R function, using default parameterization (black squares and segments) or optimized MCMC parameterization (surperimposed gray squares and segments). Squares represent the posterior mean of $\log{(HR)}$ and segments the 95\% credibility interval for Bayesian methods, the maximum likelihood point estimate and 95\% confidence interval for frequentist methods. $^\star$ function\{package\} or reference publication.} \label{forest_plot_sensi}
\end{figure}

Table \ref{tab:loghr3} presents the estimates obtained on the data derived from the ALLOZITHRO trial, with optimized and homogeneous parameterization for functions using MCMC, and figure \ref{forest_plot_sensi} illustrates the comparison between estimates with default parameterization and those optimized (as relevant for functions using MCMC).

\begin{table}[!h]
    \caption{Sensitivity analyses with optimized MCMC simulation: optimized MCMC parameterizations in the R packages for Bayesian survival PH models (number of chains $=4$ and burn-in to total number of iterations ratio $=0.5$ for all), PEM: piecewise exponential model.}
    \label{tab:mcmcparameters}
\begin{footnotesize}
\centering
    \begin{tabular}{lcc}
    \hline
     
      Package/Method  & Total no. of iterations & Thin \\
        \hline
        \multicolumn{3}{l}{\underline{\texttt{SemiCompRisks} package}}\\
       \quad \emph{BayesSurv$\_$HReg(, model='Weibull')}& $4.10^5$ & 200  \\
    \quad \emph{BayesSurv$\_$HReg(, model='PEM')}& $2.10^5$ & 100  \\
       
  \multicolumn{3}{l}{\underline{Stan via \texttt{survHE} package}}\\      
\quad Parametric exponential & 10000 &  5  \\
\quad Parametric Weibull & 10000 &  5  \\    
  \multicolumn{3}{l}{\underline{Stan via \texttt{rstanarm} package}}\\      
\quad Parametric exponential & 10000 &  5  \\
\quad Parametric Weibull & 10000 &  5  \\   
\quad M-splines & 10000 &  5  \\ 
\quad B-splines & 10000 &  5  \\ 

          \multicolumn{3}{l}{\underline{\texttt{dynsurv} package}}\\
          \quad \emph{bayesCox()} & 10000 &  5 \\
 \multicolumn{3}{l}{\underline{\texttt{spBayesSurv} package}}\\
        \quad \emph{indeptCoxph()}& 20000 & 10 \\
         \quad \emph{survregbayes(..., survmodel="PH", dist="weibull")}& 10000 & 5 \\
        \quad \emph{survregbayes(..., survmodel="PH", dist="loglogistic")}& 10000 & 5 \\
       \quad  \emph{survregbayes(..., survmodel="PH", dist="lognormal")}& 10000 & 5  \\
         \quad \emph{survregbayes2(..., survmodel="PH", dist="weibull")} & 10000 & 5  \\
        \quad  \emph{survregbayes2(..., survmodel="PH", dist="loglogistic")} & 10000 & 5 \\
        \quad  \emph{survregbayes2(..., survmodel="PH", dist="lognormal")} & 10000 & 5 \\
\underline{Murray et al (2016)} & 10000 & 5\\ 
         \hline
    \end{tabular}

\end{footnotesize}
\end{table}

\begin{table}[!h]
\begin{small}
    \centering
        \caption{ALLOZITHRO Trial: Sensitivity analysis for Bayesian functions using MCMC sampling, with optimized parameterization. Estimates of the log hazard ratio (HR) of event (death or airflow decline) in the azithromycin group over placebo according to the method. 95\%CI refers to the credible intervals. PEM: piecewise exponential model; TBP: transformed Bernstein polynomial; MPT: mixture of Polya trees}\label{tab:loghr3}
    \begin{tabular}{lccc}
    \hline
         Estimation method  & \multicolumn{2}{c}{Posterior $log HR$} &\\ \cline{2-3}
       &  Mean&95\% CI& $Pr (HR>1.5)$\\
        \hline
 		\multicolumn{4}{l}{\underline{\texttt{SemiCompRisks} package}}\\
        \quad Parametric Weibull & 0.3508  &0.0905 to 0.6098  & 0.3475\\
        \quad PEM & 0.3455 & 0.1622 to 0.5700 & 0.3100 \\
		\multicolumn{4}{l}{\underline{Stan via \texttt{survHE} package}}\\
        \quad Parametric exponential &  0.3505 & 0.0924 to 0.6144 & 0.3415\\
        \quad Parametric Weibull &  0.3470 & 0.0860 to 0.6078 &  0.3233\\        
		\multicolumn{3}{l}{\underline{Stan via \texttt{rstanarm} package}$^\star$}\\
        Parametric exponential & 0.3503 & 0.1011 to 0.6135 & 0.3313\\
        Parametric Weibull  & 0.3550 & 0.0956 to 0.6356 &0.3483\\        
				M-splines  & 0.2898 & 0.0282 to 0.5439 &0.1823\\
        B-splines  & 0.3569 & 0.0953 to 0.6286 &0.3438\\    

        \multicolumn{4}{l}{\underline{\texttt{dynsurv} package}}\\
         PEM Gamma   &  0.2646  & -0.0246 to 0.5530 & 0.1668 \\
         
		\multicolumn{4}{l}{\underline{\texttt{spBayesSurv} package}}\\
        \quad PEM \emph{indeptCoxph()}& 0.3621 & 0.1004 to 0.6083  & 0.3733\\
        \quad TBP \emph{survregbayes()}&&\\
       \quad \quad \emph{dist="weibull"}     & 0.3509 & 0.0875 to 0.6197& 0.3402\\
       \quad \quad \emph{dist="loglogistic"} &  0.3451   & 0.0867 to 0.6052  & 0.3290\\
       \quad \quad \emph{dist="lognormal"} &  0.3434 & 0.0812 to 0.6016 &0.3228\\
     
	\quad	MPT \emph{survregbayes2()} & &  \\
       \quad \quad \emph{ dist="weibull"}& 0.3517  & 0.0765 to 0.6239 &0.3522\\
       \quad  \quad \emph{dist="loglogistic"} &  0.3617  & 0.0906 to  0.6427 &0.3852\\
      \quad  \quad \emph{dist="lognormal"} &  0.3542  & 0.0670 to 0.6350 & 0.3572\\
 		\multicolumn{4}{l}{\underline{Murray \emph{et al.} (2016)}}\\
  \quad LRTP Splines & 0.2202  & -0.0404 to 0.4745 &  0.0830\\
\hline
    \end{tabular}

\end{small}
{\footnotesize $^\star$ As of 22 July 2019, \texttt{rstanarm} development version, including the \emph{stan$\_$surv()} function, downloaded from the survival branch of the package available on github at https://github.com/stan-dev/rstanarm/tree/feature/survival}

\end{table}

\end{document}